\documentclass[journal]{IEEEtran}
\usepackage[cmex10]{amsmath}
\usepackage{amssymb,amsthm,amsfonts,mathrsfs,xparse}
\usepackage{graphicx}
\usepackage{dsfont,enumitem,stackrel}
\usepackage{color}
\usepackage{mathtools,amssymb,bm,mathabx}
\usepackage{amstext}
\usepackage{array}
\usepackage{algorithmicx}
\usepackage[ruled]{algorithm}
\usepackage{algpseudocode}
\usepackage{algpascal}
\usepackage{algc}
\usepackage{cases}
\usepackage{pgfplots}
\usepackage{accents}
\usepackage{caption}
\usepackage{float}
\usepackage{cite}
\usepackage [autostyle, english = american]{csquotes}
\usepackage{hyperref}
\theoremstyle{remark}

\newcommand\ASTART{\bigskip\noindent\begin{minipage}[b]{0.5\linewidth}}
	
	\newcommand\AENDSKIP{\end{minipage}\bigskip}
\newcommand\AEND{\end{minipage}}
\ifCLASSOPTIONcaptionsoff
\usepackage[nomarkers]{endfloat}
\let\MYoriglatexcaption\caption
\renewcommand{\caption}[2][\relax]{\MYoriglatexcaption[#2]{#2}}
\fi

\def\change{black}
\theoremstyle{plain}
\newtheorem{thm}{\textbf{Theorem}}
\newtheorem{lem}{\textbf{Lemma}}
\newtheorem{prop}{\textbf{Proposition}}

\newtheorem{result}{\textbf{Result}}
\theoremstyle{definition}
\newtheorem{defn}{\textbf{Definition}}

\theoremstyle{remark}
\newtheorem{rem}{\bf Remark}
\newtheorem*{sketch}{\bf Proof skech }

\newcommand*{\rom}[1]{\expandafter\@slowromancap\romannumeral #1@}

\graphicspath{{./Figures/}} 
\begin{document}
%
\title{On the Error in Phase Transition Computations for Compressed Sensing}
\author{Sajad~Daei, Farzan~Haddadi, Arash~Amini, Martin Lotz
\thanks{S. Daei and F. Haddadi are with the School of Electrical Engineering, Iran University of Science \& Technology. A. Amini is with EE department, Sharif University of Technology. M. Lotz is with the Mathematics Institute, University of Warwick.}
}%

\maketitle

\begin{abstract}
Evaluating the statistical dimension is a common tool to determine the asymptotic phase transition in compressed sensing problems with Gaussian ensemble. Unfortunately, the exact evaluation of the statistical dimension is very difficult and it has become standard to replace it with an upper-bound. To ensure that this technique is suitable, \cite{amelunxen2013living} has introduced an upper-bound on the gap between the statistical dimension and its approximation. In this work, we first show that the error bound in \cite{amelunxen2013living} in some low-dimensional models such as total variation and $\ell_1$ analysis minimization becomes poorly large. Next, we develop a new error bound which significantly improves the estimation gap compared to \cite{amelunxen2013living}. In particular, unlike the bound in \cite{amelunxen2013living} that fails in some settings with overcomplete dictionaries, our bound exhibits a decaying behavior in such cases.
\end{abstract}

\begin{IEEEkeywords}
statistical dimension, error estimate, low-complexity models.
\end{IEEEkeywords}

%
\IEEEpeerreviewmaketitle

\section{Introduction}
 \IEEEPARstart{U}{nderstanding} the behavior of random compressed sensing problems in transition from absolute failure to success (known as phase transition) has been the subject of research in recent years
 \cite{chandrasekaran2012convex,amelunxen2013living,donoho2009counting,rudelson2008sparse,bayati2015universality,stojnic2009various, donoho2013accurate, oymak2016sharp}. Most of these works concentrate on simple sparse models and do not allude to the challenges in other low-dimensional structures such as low-rank matrices, block-sparse vectors, gradient-sparse vectors and cosparse (also known as analysis sparse\cite{elad2007analysis}) vectors. For simplicity, we associate such structures with their common recovery techniques and rename the structures accordingly. For instance, total variation (TV), $\ell_1$ analysis and $\ell_{1,2}$ minimization refer to both the recovery techniques and the underlying low-dimensional structures. In this work, we revisit linear inverse problems with the aim of  recovering a vector $\bm{x}\in\mathbb{R}^n$ from a few random linear measurements $\bm{y}=\bm{A}\bm{x}\in\mathbb{R}^m$. This is summarized as solving the following convex program:
 \begin{align}
 \mathsf{P}_f: ~~&\min_{\bm{z}\in \mathbb{R}^n} f(\bm{z})\nonumber\\
 &\mathrm{s.t.}~~\bm{y}=\bm{A}\bm{z},
 \end{align}  
where, $\bm{A}\in\mathbb{R}^{m\times n}$ is the measurement matrix whose entries are i.i.d. random variables with normal distribution and $f$ is a convex penalty function that promotes the low-dimensional structure. A major subject of recent research is the number of Gaussian measurements (the number of rows in $\bm{A}$) one needs to recover a structured vector $\bm{x}$ from $\bm{y}\in\mathbb{R}^m$. In \cite{donoho2009counting}, a bound is obtained using polytope angle calculations with asymptotic sharpness in case of $f=\|\cdot\|_1$. A link between the number of required measurements and the error in the denoising problem is investigated in \cite{oymak2016sharp}. For the particular case of $f=\|\cdot\|_{\mathrm{TV}}$ in $\mathsf{P}_f$, we need to consider the denoising problem
\begin{align}
\widehat{\bm{x}}=\mathop{\arg\min}_{\bm{z}\in \mathbb{R}^n}\tau \sigma\|\bm{z}\|_{\mathrm{TV}}+\frac{1}{2}\|\bm{y}-\bm{z}\|_2^2,
\end{align}
where $\bm{y}=\bm{x}+\bm{w}$ with $\bm{w}$ being an additive noise drawn from $\mathcal{N}(\bm{0},\sigma^2\bm{I})$. If we define the worst-case normalized mean squared error (NMSE) as 
\begin{align}
\text{NMSE}:=\lim\limits_{\sigma\rightarrow 0}\inf_{\tau\ge0}\frac{\mathds{E}\|\bm{x}-\widehat{\bm{x}}\|_2^2}{\sigma^2},
\end{align}
then, the result in \cite{oymak2016sharp} implies that the value of NMSE is a sharp estimate of the required number of measurements (for $\mathsf{P}_{\rm TV}$) in the asymptotic case.
%
Also, in \cite{donoho2013accurate}, the authors showed that the mentioned NMSE is the same as the number of required measurements that TV approximate message passing (TV-AMP) algorithm needs. \cite{chandrasekaran2012convex} introduced a general framework for obtaining the number of Gaussian measurements in different low-dimensional structures using Gordon min-max inequality\cite{gordon1988milman} and the concept of atomic norms. Specifically, it was shown that $\omega^2(\mathcal{D}(f,\bm{x})\cap \mathbb{B}^n)+1$ measurements are sufficient. Here, $\mathcal{D}(f,\bm{x})$ is the descent cone of $f$ at $\bm{x}\in\mathbb{R}^n$ and $\omega^2(\mathcal{D}(f,\bm{x})\cap \mathbb{B}^n)$ is the squared Gaussian width, which intuitively measures the size of this cone. In \cite{amelunxen2013living}, it has been shown that the statistical dimension of this cone, which is defined below and differs from the squared Gaussian width above by at most $1$, specifies the phase transition of the [random] convex program $\mathsf{P}_f$ from absolute failure to absolute success:
\begin{align}
\delta(\mathcal{D}(f,\bm{x})):=\mathds{E}~\mathrm{dist}^2(\bm{g},\mathrm{cone}(\partial f(\bm{x}))).
\end{align}
$\delta(\mathcal{D}(f,\bm{x}))$ is the average distance of a standard Gaussian i.i.d. vector $\bm{g}\in\mathbb{R}^n$ from non-negative scalings of the subdifferential at point $\bm{x}\in\mathbb{R}^n$. So far, we know that a phase transition exists in $\mathsf{P}_f$ and its boundary is interpreted via the statistical dimension. A natural question is how we can find an expression for the phase transition curve. The upper-bound for $\delta(\mathcal{D}(f,\bm{x}))$, first used in the context of $\ell_1$ minimization by Stojnic ( \cite{stojnic2009various}), is given by:
\begin{align}\label{eq.upperbound}
&\delta(\mathcal{D}(f,\bm{x}))\le\inf_{t\ge0}\mathds{E}~\mathrm{dist}^2(\bm{g},t\partial f(\bm{x})):=U_{\delta}.
\end{align}
However, it is still unknown whether $U_\delta$ is sharp for different low-dimensional structures. For ease of notation, we define the error $E_{\delta}$ by:
\begin{align}
&E_{\delta}:=U_{\delta}-\delta(\mathcal{D}(f,\bm{x}))\label{eq.error_delta}.
\end{align}
Here, $U_{\delta}$ represents a sufficient number of measurements that $\mathsf{P}_f$ needs for successful recovery. In \cite{amelunxen2013living}, implicit formulas are derived for the upper-bound (\ref{eq.upperbound}) in case of $\ell_1$ and nuclear norm. Recently, an explicit upper-bound for $U_{\delta}$ in case of $\ell_1$ analysis and TV minimization is presented in \cite{genzel2017ell}. The proposed bound depends on a notion called \lq\lq generalized analysis sparsity\rq\rq~ and is numerically observed to be tight for many analysis operators. 
{\color{\change}In \cite{amelunxen2013living}, a general upper-bound expression for $E_{\delta}$ (known as the error estimate) for various structure-inducing functions (e.g. $\ell_1$) is proposed (see Theorem \ref{thm.amelunxen}). For the cases of $\ell_1$ and nuclear norm minimization, it is further shown that the normalized error estimate $\frac{\widehat{E}_p}{n}$ (and thus $\frac{E_{\delta}}{n}$) vanishes in high dimensions. This result confirms that $U_{\delta}$ is a good surrogate for $\delta(\mathcal{D}(f,\bm{x}))$. However, the asymptotic behavior of $\frac{E_{\delta}}{n}$ in cases of}
$\|\cdot\|_{1,2}$, $\|\bm{\Omega}\cdot\|_1$ and $\|\cdot\|_{\mathrm{TV}}:=\|\bm{\Omega_d}\cdot\|_1$ where
\begin{align}
\bm{\Omega}_d=\begin{bmatrix}
1& -1 & 0 & \cdots & 0 \\
0& 1 & -1 & \cdots & 0  \\
&  \ddots    & \ddots       & \ddots & \\
0 & \cdots & \cdots & 1 & -1 &
\end{bmatrix}\in\mathbb{R}^{n-1\times n},
\end{align}
{\color{\change} is not studied. Thus, one could simply think of the following problems:}
\begin{enumerate}
	\item \label{item.1} Does $U_{\delta}$ provide a fair estimate of the statistical dimension?
	\item \label{item.2}How to quantify the gap between the exact phase transition curve and the one obtained via $U_{\delta}$?
	\item \label{item.3}Can one extend the previous error bounds obtained for $\ell_1$ minimization in \cite{amelunxen2013living} to other low-dimensional structures such as block sparsity, TV and $\ell_1$ analysis?
\end{enumerate}

In this work, we try to find answers to these questions. Specifically, we want to study how well $U_{\delta}$ describes $\delta(\mathcal{D}(f,\bm{x}))$ in low-dimensional structures represented by $\|\cdot\|_1$, $\|\cdot\|_{1,2}$, $\|\cdot\|_*$, $\|\bm{\Omega}\cdot\|_1$ and $\|\cdot\|_{\mathrm{TV}}:=\|\bm{\Omega_d}\cdot\|_1$. 
A generic $\bm{\Omega}\in\mathbb{R}^{p\times n}$ can be tall, i.e. $p> n$, or fat, i.e. $p\le n$; due to the similarity of the arguments used in this paper, we include square matrices in the category of fat matrices.
Tall $\bm{\Omega}$ matrices cover various redundant\footnote{The term redundant refers to an analysis operator with more number of rows than columns.} analysis operators that are common in practice; in particular, redundant wavelet frames\cite{casazza2013introduction}, and redundant random frames (widely used as a benchmark template in \cite{kabanava2015analysis,nam2013cosparse,genzel2017ell,elad2007analysis}). Also fat matrices  include  examples such as the one-dimensional finite difference operator $\bm{\Omega}_d$ and non-redundant random analysis operators (used in \cite[Section 3.3]{genzel2017ell}). 
We call a signal $\bm{x}\in\mathbb{R}^n$ an analysis-sparse vector (also called as cosparse vector \cite{elad2007analysis}) with respect to the analysis operator $\bm{\Omega}\in\mathbb{R}^{p\times n}$ if after applying $\bm{\Omega}$ the resulting vector becomes sparse; i.e., $\bm{\Omega x}$ is sparse. 
We denote the support set of $\bm{\Omega x}$ with $\mathcal{S}$; similarly, $\overline{\mathcal{S}}$ stands for the zero set of $\bm{\Omega x}$. 
The number of zeros in $\bm{\Omega x}$, i.e., $|\overline{\mathcal{S}}|$, is called the cosparsity of $\bm{x}$ with respect to $\bm{\Omega}$\cite{nam2013cosparse,elad2007analysis,genzel2017ell}. 
{\color{\change} We should highlight that in most of the existing literature regarding the $\ell_1$-analysis problem it is assumed that  $\bm{\Omega}\in\mathbb{R}^{p\times n}$  has rows in general position\footnote{Every subset of $n$ rows of $\bm{\Omega}$ are linearly independent.}}.
\subsection{Motivation}
Tables \ref{table.l1} and \ref{table.TV} present the results of a computer experiment designed to evaluate the error of $U_{\delta}$ in estimating the statistical dimension. In two experiments shown in Tables \ref{table.l1} and \ref{table.TV}, we test the error bound for $\ell_1$ and TV minimization. The value of $\delta(\mathcal{D}(f,\bm{x}))$ which is approximately equal to $\omega^2(\mathcal{D}(f,\bm{x})\cap\mathbb{B}^n)$, is computed using the procedure proposed in \cite[Section B.2]{genzel2017ell} (see Appendix \ref{sec.TrueGaussianwidth}). 
In the first experiment, 
for each sparsity level, we construct a sparse vector $\bm{x}\in\mathbb{R}^{1000}$ with random non-zero values (distributed as $\mathcal{N}(\bm{0},\sqrt{1000})$) at uniformly random locations. 
In the second experiment, we set $\bm{\Omega}=\bm{\Omega}_d$, and generate a gradient sparse vector $\bm{x}\in\mathbb{R}^{1000}$ as the sum of two components: a small but non-zero component in the space ${\rm null}({\bm{\Omega}}_{\overline{S}})$, and a large component in the space ${\rm null}({\bm{\Omega}})$. The notation ${\bm{\Omega}}_{\overline{\mathcal{S}}}$ refers to the matrix $\bm{\Omega}$ restricted to the rows indexed by $\overline{\mathcal{S}}$ (see Section \ref{sec.ame_fail} for more details).
%
For Tables \ref{table.l1} and \ref{table.TV}, the upper-bound (\ref{eq.upperbound}) is obtained by \cite[Equation D.6]{amelunxen2013living} and numerical optimization\footnote{See also \cite[Section 4]{zhang2016precise} which proposes a numerical method to calculate $U_{\delta}$ in case of TV minimization.}, respectively. {\color{\change} As shown in Tables \ref{table.l1} and \ref{table.TV}, there exists a gap between the true error $E_{\delta}$ and the state of the art theoretical error estimate $\widehat{E}_p$ in (\ref{eq.errorbound}).  While the normalized gap, i.e., $\frac{|E_{\delta}-\widehat{E}_p|}{n}$ is negligible in the $\ell_1$ case (Table \ref{table.l1}), it is considerable in case of TV minimization (Table \ref{table.TV})}. Now, a natural question that arises is: can we find a better bound that reduces the gap?
\begin{table}[t]
\centering
{	\color{\change}
	
		\begin{tabular}{ |c|c|c|c|l|}
			\hline
			$U_{\delta}$&$\delta(\mathcal{D}(\|\cdot\|_1,\bm{x}))$&$\frac{E_{\delta}}{n}$ & $\frac{\widehat{E}_p}{n}$ in (\ref{eq.errorbound}) & $s$\\ 
			\hline
		$9.458$&$9.0905$&$0.0003$ & $0.064$ & $1$ \\
			\hline
			$16.828$&$16.8$&$0.0003$ & $0.045$ & $2$\\ 
			\hline
			$23.4544$&$23.04$&$.0005$ & $0.036$ & $3$ \\
			\hline
			$61.244$&$60.84$&$0.0004$ & $0.02$ & $10$ \\
			\hline
			$104.1814$&$104.04$&$0.0001$ & $0.015$ & $20$ \\
\hline
	\end{tabular}
}
\caption{In this table, we examine the error estimate \cite{amelunxen2013living} for function $f=\|\cdot\|_1$ and signal $\bm{x}$ with dimension $n=1000$. The upper-bound $U_{\delta}$ is computed using the package SNOW in \cite{amelunxen2013living}. The true sample complexity is denoted by $\delta(\mathcal{D}(\|\cdot\|_1,\bm{x}))$ and computed by squaring $\omega(\mathcal{D}(\|\cdot\|_1,\bm{x})\cap \mathbb{B}^n)$ (see Appendix \ref{sec.TrueGaussianwidth} for more details). For the considered class of $\bm{x}$ signals, {\color{\change}we observe that there is a negligible gap between
 $\frac{\widehat{E}_p}{n}$ in \cite{amelunxen2013living} and the true normalized error $\frac{E_{\delta}}{n}$.}}\label{table.l1}
\end{table} 
\begin{table*}[t]
	\centering
	{\color{\change}
	\begin{tabular}{ |c|c|c|c|c|c|c|}
		\hline
	$U_{\delta}$&$\delta(\mathcal{D}(\|\cdot\|_{\rm TV},\bm{x}))$&$\frac{E_{\delta}}{n}$ & $\frac{\widehat{E}_p}{n}$ in (\ref{eq.errorbound}) & Average amplitude of $\bm{x}$&$\|\bm{\Omega x}\|_{\infty}$&$s$\\ 
		\hline
		$58$&$56.25$&$0.002$ & $2063.3$ & $1000$&$0.1$&$10$ \\
		\hline
		$61.1$&$60.84$&$0.0013$ & $5.1143$ & $30$&$4$&$10$\\ 
		\hline
		$80.03$&$79.21$&$0.0008$ & $0.4$ & $3$&$0.9$&$10$ \\
		\hline
	$982$&$979.69$&$0.002$ & $10.4$ & $316$&$0.4$&$850$ \\
		\hline
	\end{tabular}
}
\caption{In this table, we examine the TV error bound for gradient sparse signals with dimension $n=1000$. The upper-bound $U_{\delta}$ is obtained by Monte Carlo simulations and numerical optimization. The true sample complexity $\delta(\mathcal{D}(\|\cdot\|_{\rm TV},\bm{x}))\approx \omega^2(\mathcal{D}(\|\cdot\|_{\rm TV},\bm{x})\cap\mathbb{B}^n)$ is obtained using the approach presented in Appendix \ref{sec.TrueGaussianwidth}. {\color{\change}The large values of $\frac{\widehat{E}_p}{n}$ observed in this table are not necessarily caused by approximating $\delta(\mathcal{D}(f,\bm{x}))$ with  $U_{\delta}$.}} \label{table.TV}
\end{table*}

\subsection{Contributions}
In this work, we rigorously analyze the error of estimating the phase transition. The significance of this error is to have a good understanding about the required number of measurements that $\mathsf{P}_f$ needs to recover a structured vector from under-sampled measurements. Our analysis is general and holds for a variety of low-dimensional structures including sparse, block-sparse, analysis sparse and gradient-sparse vectors, as well as low-rank matrices. In brief, the contributions of this work can be listed as follows.

\begin{enumerate}
	\item  {\itshape{Identifying a failure regime for \cite{amelunxen2013living}}}: For $f=\|\bm{\Omega}\cdot\|_1$ in $\mathsf{P}_f$, the error estimate of \cite{amelunxen2013living} shown in \eqref{eq.errorbound}, can become remarkably large for some specific signals and analysis operators. For fat analysis operators, a typical signal $\bm{x}$ with such property is constructed as:
	\begin{align}
	\bm{x}=\bm{P}_{{\rm null}(\bm{\Omega}_{\overline{\mathcal{S}}})} \bm{w}+\bm{P}_{{\rm null}(\bm{\Omega})}\bm{c}\in\mathbb{R}^n,
	\end{align}
	where $\bm{w},\bm{c}\in\mathbb{R}^n$ are arbitrary vectors. For tall analysis operators, we can find pairs of $\bm{\Omega}$ and $\bm{x}$ for which the error estimate \cite{amelunxen2013living} explodes.
We precisely investigate this in Section \ref{section.priorworkeva}.

	\item {\textit{Obtaining an error bound for $\delta(\mathcal{D}(f,\bm{x}))$ with rather general $f(\cdot)$}}: $\delta(\mathcal{D}(f,\bm{x}))$ precisely determines the boundary of failure and success of $\mathsf{P}_f$. However, exact computation of $\delta(\mathcal{D}(f,\bm{x}))$ is very difficult. It is common to approximate $\delta(\mathcal{D}(f,\bm{x}))$ with $U_{\delta}$. By providing an error bound, we formally show that this approximation is good. More precisely, we show that
	\begin{align}
\frac{E_{\delta}}{n}\le h_2(\beta,\omega),
	\end{align}   
where $\beta$ depends on $\partial f(\bm{x})$, and $h_2(\beta,\omega)$ is a function of $\beta$ and $\omega(\mathcal{D}(f,\bm{x})\cap \mathbb{B}^n)$ that is succinctly shown by $\omega$. Under certain conditions, we show that 
$h_2(\beta,\omega)$ vanishes as $n$ grows sufficiently large. To a great extent, the setting considered for $f$ (see (\ref{eq.mycondition})) is nonrestrictive. In particular, it includes the important special cases of $\|\bm{\Omega}\cdot\|_1$ for tall and fat analysis operators. 

In contrast to the error estimate of \cite{amelunxen2013living} that directly depends on $\bm{x}\in\mathbb{R}^n$, our bound is determined by $\partial f(\bm{x})$. Besides, our error bound holds even for rank-deficient fat $\bm{\Omega}\in\mathbb{R}^{p\times n}$ matrices. {\color{\change}We should emphasize that our error estimate bound is not sharp in all cases of analysis operators and does not necessarily fill the gap between $E_{\delta}$ and $\widehat{E}_p$ in \eqref{eq.errorbound}. In fact, there are various settings in which the bound in \cite{amelunxen2013living}, our bound, or both are effective.  
}


%
\end{enumerate}

\subsection{Notation}
Throughout the paper, scalars are denoted by lowercase letters, vectors by lowercase boldface letters, and matrices by uppercase boldface letters. The $i$th element of the vector $\bm{x}$ is given either by ${x}(i)$ or $x_i$. The notation $(\cdot)^\dagger$ stands for the pseudo-inverse operator. We reserve calligraphic uppercase letters for sets (e.g. $\mathcal{S}$) and denote the cardinality of a set $\mathcal{S}$ by $|\mathcal{S}|$. The complement of a set $\mathcal{S}$ in $\{1,..., n\}$ (briefly represented as $[n]$) is denoted by ${\overline{\mathcal{S}}}$.
	Similarly, the complement of an event $\mathcal{E}$ is shown by $\mathcal{\overline{E}}$. For a matrix $\bm{X}\in\mathbb{R}^{m\times n}$ and a subset $\mathcal{S}\subseteq [n]$, the notation $\bm{X}_\mathcal{S}$ refers to the sub-matrix of $\bm{X}$ by including the rows indexed by $\mathcal{S}$. Similarly, for $\bm{x}\in\mathbb{R}^n$, $\bm{x}_\mathcal{S}$ stands for the vector in $\mathbb{R}^{n}$ that coincides with $\bm{x}$ at entries indexed by $\mathcal{S}$ and zero elsewhere. Also, we use the notation $\widetilde{\bm{x}}_{\mathcal{S}}$ to represent a sub-vector of $\bm{x}$ in $\mathbb{R}^{|\mathcal{S}|}$, that is formed by discarding the zero entries not indexed in $\mathcal{S}$.
The null-space of linear operators is denoted by $\mathrm{null}(\cdot)$. For a matrix $\bm{\Omega}$, the operator norm is defined as $\|\bm{\Omega}\|_{p\rightarrow q}=\underset{\|\bm{x}\|_p\le1}{\sup}\|\bm{\Omega x}\|_q$. Also, {\color{\change}$\kappa(\bm{\Omega}):=\frac{\sigma_{\max}(\bm{\Omega})}{\sigma_{\min}(\bm{\Omega})}$ denotes the condition number of $\bm{\Omega}$}. The polar $\mathcal{K}^{\circ}$ of a cone $\mathcal{K}\subset\mathbb{R}^n$ is the set of vectors forming non-acute angles with every vector in $\mathcal{K}$, i.e. \begin{align}
\mathcal{K}^\circ=\{\bm{v}\in\mathbb{R}^n: \langle \bm{v}, \bm{z} \rangle\le 0~\forall \bm{z}\in\mathcal{K}\}.
\end{align}
{\color{\change}$\mathds{B}^n$ and $\mathbb{S}^{n-1}$ stand for the unit ball $\{\bm{x}\in \mathbb{R}^n:~\|\bm{x}\|_2\le 1\}$ and unit sphere $\{\bm{x}\in \mathbb{R}^n:~\|\bm{x}\|_2= 1\}$, respectively.}
$\bm{P}_{\mathcal{C}}$ is the matrix associated with the orthogonal projection  onto the subspace $\mathcal{C}$, that maps a vector in $\mathbb{R}^n$ onto the subspace $\mathcal{C}\subset \mathbb{R}^n$.

\subsection{Outline}
The paper is organized as follows. 
The required concepts from convex geometry are reviewed in Section \ref{section.convexgeometry}. Section \ref{section.related} discusses two approaches in obtaining the error estimate. Section \ref{section.mainresult} is dedicated to present our main contributions. In Section \ref{section.priorworkeva}, we investigate the estimate in \cite{amelunxen2013living} and introduce some examples for which the error estimate does not work. In Section \ref{section.simulation}, numerical experiments are presented which confirm our theory. Finally, the paper is concluded in Section \ref{section.conclusion}.

\section{Convex Geometry}\label{section.convexgeometry}
In this section, a review of basic concepts of convex geometry is provided.
\subsection{Descent Cones}
The descent cone $\mathcal{D}(f,\bm{x})$ at a point $\bm{x}\in\mathbb{R}^n$ consists of the set of directions that do not increase $f$ and is given by:
\begin{align}\label{eq.descent cone}
\mathcal{D}(f,\bm{x})=\bigcup_{t\ge0}\{\bm{z}\in\mathbb{R}^n: f(\bm{x}+t\bm{z})\le f(\bm{x})\}.
\end{align}
The descent cone reveals the local behavior of $f$ near $\bm{x}$ and is a convex set. There is also a relation between decent cone and subdifferential \cite[Chapter 23]{rockafellar2015convex} given by:
\begin{align}\label{eq.D(f,x)}
\mathcal{D}^{\circ}(f,\bm{x})=\mathrm{cone}(\partial f(\bm{x})):=\bigcup_{t\ge0}t\partial f(\bm{x}).
\end{align}
\subsection{Statistical Dimension}
\begin{defn}{Statistical Dimension}\cite{amelunxen2013living}:
	Let $\mathcal{C}\subseteq\mathbb{R}^n$ be a closed convex cone. The statistical dimension of $\mathcal{C}$ is defined as:
	\begin{align}\label{eq.statisticaldimension}
	\delta(\mathcal{C}):=\mathds{E}\|\mathcal{P}_\mathcal{C}(\bm{g})\|_2^2=\mathds{E}~\mathrm{dist}^2(\bm{g},\mathcal{C}^\circ),
	\end{align}
	where $\mathcal{P}_\mathcal{C}(\bm{x})$ is the projection of $\bm{x}\in \mathbb{R}^n$ onto the set $\mathcal{C}$ defined as: $\mathcal{P}_\mathcal{C}(\bm{x})=\underset{\bm{z} \in \mathcal{C}}{\arg\min}~\|\bm{z}-\bm{x}\|_2$.
\end{defn}
The statistical dimension extends the concept of linear subspaces to convex cones. Intuitively, it measures the size of a cone. Furthermore, $\delta(\mathcal{D}(f,\bm{x}))$ determines the precise location of transition from failure to success in $\mathsf{P}_f$.

\subsection{Gaussian width}
\begin{defn}
	The Gaussian width of a set $C$ is defined as:
	\begin{align}
	\omega(C):=\mathds{E}\sup_{\bm{y}\in C}\langle \bm{y},\bm{g}\rangle.
	\end{align}
\end{defn}
The relation between statistical dimension and Gaussian width is summarized in the following \cite[Proposition 3.6]{chandrasekaran2012convex},\cite[Proposition 10.2]{amelunxen2013living}.
\begin{align}
&\omega(C\cap\mathbb{S}^{n-1})\le\omega(C\cap\mathbb{B}^n)=\mathds{E}\|\mathcal{P}_{C}(\bm{g})\|_2=\mathds{E}~\mathrm{dist}(\bm{g},C^\circ),
\end{align}
\begin{align}\label{eq.lazem}
\omega^2(C\cap \mathbb{B}^{n})=(\mathds{E}\|\mathcal{P}_{C}(\bm{g})\|_2)^2\le \delta(C).
\end{align}
{\color{\change} It is shown in \cite[Proposition 10.2]{amelunxen2013living}. that the quantities $\omega^2(C\cap \mathbb{B}^{n})$ and $\delta(C)$} differ numerically by at most $1$.
\subsection{Optimality Condition}
In the following, we characterize when $\mathsf{P}_f$ succeeds in the noise-free case.
\begin{prop}\cite[Proposition 2.1]{chandrasekaran2012convex} Optimality condition: Let $f$ be a proper convex function. The vector $\bm{x}\in \mathbb{R}^n$ is the unique optimal point of $\mathsf{P}_f$ if and only if $\mathcal{D}(f,\bm{x})\cap \mathrm{null}(\bm{A})=\{\bm{0}\}$.
\end{prop}
The next theorem determines the number of measurements needed for successful recovery of $\mathsf{P}_f$ for any proper convex function $f$.
\begin{thm}\label{thm.Pfmeasurement}\cite[Theorem 2]{amelunxen2013living}:
	Let $f:\mathbb{R}^n\rightarrow \mathbb{R}\cup \{\pm\infty\}$ be a proper convex function and $\bm{x}\in \mathbb{R}^n$ a fixed vector. Suppose that $m$ independent Gaussian linear measurements of $\bm{x}$ are observed via $\bm{y}=\bm{Ax} \in \mathbb{R}^m$. If
	\begin{align}
	m\ge \delta(\mathcal{D}(f,\bm{x}))+\sqrt{8\log(\frac{4}{\eta})n},
	\end{align}
	for a given probability of failure (tolerance) $\eta \in [0,1]$, then, we have
	\begin{align}
	\mathds{P}(\mathcal{D}(f,\bm{x})\cap \mathrm{null}(\bm{A})=\{\bm{0}\})\ge 1-\eta.
	\end{align}
	Besides, if
	\begin{align}
	m\le \delta(\mathcal{D}(f,\bm{x}))-\sqrt{8\log(\frac{4}{\eta})n},
	\end{align}
	 then,
	 \begin{align}
	 \mathds{P}(\mathcal{D}(f,\bm{x})\cap \mathrm{null}(\bm{A})=\{\bm{0}\})\le \eta.
	 \end{align}
\end{thm}

\section{Related Works in Error Estimation}\label{section.related}
For bounding the distance between $\delta(\mathcal{D}(f,\bm{x}))$ and $U_{\delta}$, two different approaches are proposed in \cite{amelunxen2013living,foygel2014corrupted}. In the following, we briefly describe these methods.
\begin{result}\cite[Theorem 4.3]{amelunxen2013living}\label{thm.amelunxen} Let $f$ be a norm. Then, for any $\bm{x}\in \mathbb{R}^n\setminus\{\bm{0}\}$:
	\begin{align}\label{eq.errorbound}
	0\le E_{\delta}\le\frac{ \color{\change} \overbrace{\color{black} 2\sup_{s\in \partial f(\bm{x})}\|s\|_2}^{\rm{Num}_{E}}}{f\left(\frac{\bm{x}}{\|\bm{x}\|_2}\right)}:=\widehat{E}_p.
	\end{align}
\end{result}   

\begin{result}
\label{prop.foygel}\cite[Proposition 1]{foygel2014corrupted}
Suppose that for $\bm{x}\in\mathbb{R}^n\setminus \{\bm{0}\}$, $\partial f(\bm{x})$ satisfies a weak decomposability assumption:
\begin{align}\label{eq.weakdecom}
\exists \bm{z}_0\in\partial f(\bm{x})~ \mathrm{s.t.}~ \langle \bm{z}-\bm{z}_0,\bm{z}_0 \rangle=0,~~ \forall \bm{z}\in \partial f(\bm{x}).
\end{align}	
Then\footnote{{\color{\change}This result is wrongly interpreted as $\inf_{t\ge0}\sqrt{\mathds{E}~\mathrm{dist}^2(\bm{g},t\partial f(\bm{x}))}\le \omega(\mathcal{D}(f,\bm{x})\cap \mathbb{B}^n)+6$ in \cite{foygel2014corrupted}.}},
\begin{align}\label{eq.errorGaussian}
{\color{\change} 
\inf_{t\ge0}\mathds{E}~\mathrm{dist}(\bm{g},t\partial f(\bm{x})) \le \omega(\mathcal{D}(f,\bm{x})\cap \mathbb{B}^n)+6.
}
\end{align}	
\end{result}

\subsection{Explanations}
Result \ref{prop.foygel} presents an error estimate for the Gaussian width of the descent cone (restricted to the unit ball) that is used to upper-bound the number of Gaussian measurements in various low-dimensional structures \cite[Section 3.1]{chandrasekaran2012convex}. For functions $f=\|\cdot\|_1$, $\|\cdot\|_{1,2}$ and $\|\cdot\|_*$, the constraint (\ref{eq.weakdecom}) is satisfied (see Section \ref{sec.betacomputations}). For $f=\|\bm{\Omega} \cdot\|_1$, however, this constraint is not generally guaranteed (see Appendix \ref{sec.l1ana_fail}).

Unlike Result \ref{prop.foygel}, the error estimate (\ref{eq.errorbound}) depends on $\partial f$ at the ground-truth vector $\bm{x}$, and the vector $\bm{x}$ itself. Although \cite[Theorem 4.3]{amelunxen2013living} restricts $f$ to be a norm, the provided proof remains valid for semi-norms such as TV.
{\color{\change}The error bound in \eqref{eq.errorbound}, is effective for many structure-inducing functions including $\ell_1$, $\ell_{1,2}$, and nuclear norm. Particularly, $\frac{\widehat{E}_p}{n}$ asymptotically vanishes in these cases. However, the normalized error estimate $\frac{\widehat{E}_p}{n}$ is large in some cases of $\ell_1$ analysis and TV minimization and does not reflect the actual error $\frac{E_{\delta}}{n}$}; in Section \ref{sec.ame_fail}, we study some examples. A naive interpretation of this fact is that $U_{\delta}$ is a poor approximation of $\delta(\mathcal{D}(f,\bm{x}))$ in those cases. Fortunately, as we show in Section \ref{section.mainresult}, this argument is invalid, which in turn suggests that (\ref{eq.errorbound}) is a loose bound in those cases.

\section{\color{\change}The study of  existing results}
\label{section.priorworkeva}

	\subsection{\color{\change}Result \ref{thm.amelunxen} for various low-complexity models}\label{sec.ame_fail}

Before we describe our contributions in Section \ref{section.mainresult}, we first evaluate the error estimate (\ref{eq.errorbound}) when $f$ is any of $\ell_1$, $\ell_{1,2}$, nuclear norm, or $\ell_1$-analysis for different analysis operators. An important observation is that the error estimate (\ref{eq.errorbound}) is increasing with 
$\tfrac{\|\bm{x}\|_2}{f(\bm{x})}$; thus, whenever this term becomes large, we might obtain a loose upper-bound. To better clarify this point, 
we study the case of $\ell_1$-analysis in three categories of fat analysis operators $\bm{\Omega}$, rank-deficient tall analysis operators $\bm{\Omega}$, and full-rank tall analysis operators $\bm{\Omega}$.

\begin{itemize}
	\item {\bf Sparse vectors}
	
Since $\delta\big(\mathcal{D}(\|\cdot\|_1,\bm{x})\big) = \delta\big(\mathcal{D}(\|\cdot\|_1,{\rm sgn}(\bm{x}))\big)$, the bound $\widehat{E}_p$ in \eqref{eq.errorbound} can be written in terms of ${\rm sgn}(\bm{x})$:
\begin{align}\label{eq.ell1}
\widehat{E}_p&=
\frac{2\sup_{\|{\bm z}_{\overline{\mathcal{S}}}\|_{\infty}\le 1}\|{\rm sgn}(\bm{x})+\bm{z}_{\overline{\mathcal{S}}}\|_2}{ \frac{\|{\rm sgn}(\bm{x})\|_1}{\|{\rm sgn}(\bm{x})\|_2} } \nonumber\\
&\ge \frac{2\|{\rm sgn}(\bm{x})+\bm{1}_{\overline{\mathcal{S}}}\|_2}{\sqrt{s}} =\frac{2\sqrt{n}}{\sqrt{s}},
%
%
%
\end{align} 
%
where the inequality is for choosing a point $\bm{1}_{\overline{\mathcal{S}}}$ in the feasible set $\|\bm{z}_{\overline{\mathcal{S}}}\|_{\infty}\le 1$. The above expression may lead to large errors in low sparsity regimes ($s\ll n$), {\color{\change} however, $\lim_{n\to \infty}\frac{\widehat{E}_p}{n}=0$. This shows that $U_{\delta}$ is asymptotically a fair approximation of $\delta\big(\mathcal{D}(\|\cdot\|_1,\bm{x})\big)$.}

	\item {\bf Block-sparse Vectors}
	
	With the same approach as in the previous case, the actual error estimate is lower-bounded by
	\begin{align}\label{eq.ell12}
	\widehat{E}_p\ge \frac{2\sqrt{q}}{\sqrt{s}} 
	\end{align}
	where $s$ and $q$ stand for the number of non-zero blocks and the total number of blocks respectively.
	Again for small $s$, the error can become large {\color{\change} while $\lim_{q\to \infty}\frac{\widehat{E}_p}{q}=0$; thus, $U_{\delta}$ is asymptotically a fair approximation of $\delta\big(\mathcal{D}(\|\cdot\|_{1,2},\bm{x})\big)$.}

	\item {\bf Low-rank Matrices} 
	
Let $\bm{X}\in\mathbb{R}^{n_1\times n_2}$ be the rank $r$  ground-truth matrix ($n_1\geq n_2$) with the  SVD decomposition $\bm{X} = \bm{U}_{n_1\times n_1} \bm{\Sigma}_{n_1\times n_2}\bm{V}_{n_2\times n_2}^H$ (alternatively, we have the reduced SVD decomposition as $\bm{X} = \bm{U}_{n_1\times r} \bm{\Sigma}_{r\times r}\bm{V}_{n_2\times r}^H$).
 Since $\delta\big(\mathcal{D}(\|\cdot\|_*,\bm{X})\big) = \delta\big(\mathcal{D}(\|\cdot\|_*,\bm{U}_{n_1\times r} \bm{V}_{n_2\times r}^H)\big) $, we can replace $\bm{X}$ with $\bm{U}_{n_1\times r}\bm{V}_{n_2\times r}^H$ in \eqref{eq.errorbound}:
%
\begin{align}\label{eq.nuclear1}
&\widehat{E}_p=\frac{2\sup_{\|\mathcal{P}_{T^\perp}(\bm{Z})\|_{2\rightarrow 2}\le 1}\|\bm{U}_{n_1\times r}\bm{V}_{n_2\times r}^H+\mathcal{P}_{T^\perp}(\bm{Z})\|_F}{ \frac{\|\bm{U}_{n_1\times r}\bm{V}_{n_2\times r}^H\|_*}{\|\bm{U}_{n_1\times r}\bm{V}_{n_2\times r}^H\|_F} },
\end{align}
where
\begin{align*}
\mathcal{P}_{T^\perp}(\bm{Z}) =  \Big(\bm{I}-\bm{P}_{{\rm span} (\bm{U}_{n_1\times r})} \Big)\bm{Z}  \Big(\bm{I}-\bm{P}_{{\rm span} (\bm{V}_{n_2\times r})} \Big).
\end{align*}
Now, by setting
\begin{align*}
\bm{Z} = \bm{U}_{n_1\times n_2}
	\begin{bmatrix}
	\bm{0}&\bm{0}\\
	\bm{0}&\bm{I}_{n_2-r}
	\end{bmatrix}
	\bm{V}_{n_2\times n_2}^H
\end{align*}
in \eqref{eq.nuclear1}, we obtain a lower-bound on $\widehat{E}_p$ as
\begin{align}\label{eq.nuclear2}
\widehat{E}_p \ge \frac{2 \|\bm{U}_{n_1\times n_2} \bm{V}_{n_2\times n_2}^H \|_{F}}{ \frac{\|\bm{U}_{n_1\times r}\bm{V}_{n_2\times r}^H\|_*}{\|\bm{U}_{n_1\times r}\bm{V}_{n_2\times r}^H\|_F} } = \frac{2\sqrt{n_2}}{\sqrt{r}}.
%
%
\end{align}
 Similar to the previous cases, when $r\ll n_2$, the bound becomes large, {\color{\change} while, $\lim_{n_1,n_2\to \infty}\frac{\widehat{E}_p}{n_1n_2}=0$. This shows that $U_{\delta}$ is asymptotically a fair approximation of $\delta\big(\mathcal{D}(\|\cdot\|_{*},\bm{x})\big)$.} 

\item {\bf Cosparse vectors (fat analysis operators)} 

For fat $\bm{\Omega}\in\mathbb{R}^{p\times n}$, the ${\rm null}(\bm{\Omega})$ is non-trivial, and we can choose $\bm{c}$ such that $\bm{P}_{{\rm null}({\bm{\Omega}})}\bm{c} \neq \bm{0}$. Now, 
if $\bm{x}=\bm{P}_{{\rm null}({\bm{\Omega}}_{\overline{\mathcal{S}}})}\bm{w}+\bm{P}_{{\rm null}({\bm{\Omega}})}\bm{c}$, where $\bm{w}$ is an arbitrary vector, the denominator of the bound in \eqref{eq.errorbound} can be written as
\begin{align}
\frac{\|\bm{\Omega x}\|_1}{\|\bm{x}\|_2} 
= 
\frac{\|{\bm{\Omega}}_{\mathcal{S}}\bm{P}_{{\rm null}({\bm{\Omega}}_{\overline{\mathcal{S}}})}\bm{w}\|_1}{\|\bm{P}_{{\rm null}({\bm{\Omega}}_{\overline{\mathcal{S}}})}\bm{w}+ \bm{P}_{{\rm null}({\bm{\Omega}})}\bm{c} \|_2}.
\end{align}
By increasing the norm of $\bm{c}$ using a scalar multiplier, the above fraction decreases. In other words, we can make the denominator of the error bound \eqref{eq.errorbound} arbitrarily small (alternatively enlarge the error bound \eqref{eq.errorbound}). One of the well-known examples in this category is the finite difference operator $\bm{\Omega}_d$, where ${\rm null}(\bm{\Omega}_d)$ consists of constant vectors. For this example, the denominator can be reduced by setting $\bm{c}=\alpha\bm{1}_{n\times 1}\in {\rm null}(\bm{\Omega}_d)$ and  $\alpha\gg 1$.
\label{item.fat_op}
\item {\bf Cosparse vectors (rank-deficient tall analysis operators)}

Similar to the previous case, ${\rm null}(\bm{\Omega})$ is non-trivial. Thus, the same approach can be devised to make the bound in \eqref{eq.errorbound} arbitrarily large.
 \label{item.rank_def_op}

\item {\bf Cosparse vectors (full-rank tall analysis operators)}

 When $\bm{\Omega}$ is a full-rank and tall matrix, we cannot generally find $\bm{x}$ that results in a small value of $\frac{\|\bm{\Omega x}\|_1}{\|\bm{x}\|_2}$. Here, we show the existence of pairs $(\bm{\Omega},\bm{x})$ for which the aforementioned ratio becomes arbitrarily small. 

Let $|\mathcal{S}|>n$, $|\overline{\mathcal{S}}|<n$, $\sigma >0$ and define
$$\bm{A}_{|\mathcal{S}|\times n} = \bm{U}_{|\mathcal{S}|\times n} \, {\rm diag}\{ \underbrace{1,1,\dots,1}_{(n-1) \text{ times}} , \sigma\} \bm{V}_{n\times n}^T,$$
where $\bm{U}_{|\mathcal{S}|\times n}$ and $\bm{V}_{n\times n}$ are arbitrary incomplete and complete unitary matrices, respectively. It is evident that $\bm{A}_{|\mathcal{S}|\times n}$ is a full-rank tall matrix. Let $\bm{u}_i$ and $\bm{v}_j$ stand for the $i$th and $j$th columns of $\bm{U}$ and $\bm{V}$, respectively. We know that $\bm{A}\,\bm{v}_n = \sigma \bm{u}_n$. Now, define $\bm{B}_{|\overline{\mathcal{S}}|\times n} = [\bm{v}_1,\dots,\bm{v}_{|\overline{\mathcal{S}}|}]^T$ (since $|\overline{\mathcal{S}}|<n$, this is possible). Obviously, $\bm{B}\,\bm{v}_n =\bm{0}$. Finally, we define
\begin{align*}
&\bm{\Omega} =\left[\begin{array}{c}
\bm{A}\\
\bm{B}
\end{array}\right],~~~~\bm{x}=\bm{v}_n,~~~~\mathcal{S}=\{1,\dots,|\mathcal{S}|\},~~~\nonumber\\ ~&\overline{\mathcal{S}}=\{|\mathcal{S}|+1,\dots,|\mathcal{S}|+|\overline{\mathcal{S}}|\}.
\end{align*}
Indeed, the design is such that $\bm{\Omega}_{\mathcal{S}}=\bm{A}$ and $\bm{\Omega}_{\overline{\mathcal{S}}}=\bm{B}$. As $\bm{A}$ has full column rank, $\bm{\Omega}$ is also a full-rank matrix. 
It is straightforward to check that $\bm{x}\in {\rm null}(\bm{\Omega}_{\overline{\mathcal{S}}})$, and
\begin{align*}
\frac{\| \bm{\Omega} \,\bm{x}\|_1}{\|\bm{x}\|_2} = \|\bm{\Omega} \,\bm{x} \|_1 = \left\| \left[\begin{array}{c}
\bm{A}\\
\bm{B}
\end{array}\right] \bm{v}_n\right\|_1= \| \bm{A} \bm{v}_n\|_1 = \sigma \|\bm{u}_n\|_1.
\end{align*}
Now, we can reduce $\sigma$ while keeping the rest untouched. In this way, we construct non-trivial pairs of $(\bm{\Omega},\bm{x})$ for which the ratio $\frac{\| \bm{\Omega} \,\bm{x}\|_1}{\|\bm{x}\|_2} $ can be set arbitrarily small.
\end{itemize}

{\color{\change}
\begin{rem}
The value of $\widehat{E}_p$ provides an upper-bound on the gap between $U_{\delta}$ and $\delta\big(\mathcal{D}(f,\bm{x})\big)$. However, its normalized value (e.g., $\frac{\widehat{E}_p}{n}$ in the $\ell_1$ minimization) is important in determining the phase transition curve. As we discussed earlier, the normalized value is vanishing in the three cases of $\ell_1$, $\ell_{1,2}$ and nuclear-norm minimization. However, we do not observe this vanishing property in some cases of  $\ell_1$-analysis. In these cases, $U_{\delta}$ might not be a good approximation of $\delta(\mathcal{D}(f,\bm{x}))$.

\end{rem}}

\begin{table*}[t]
	{\color{\change}
	\centering
	
	\begin{tabular}{|c|c|c|c|c|c|c|c|c|c|c|}
		\hline
	    Type of anal. operator  $\bm{\Omega}$&$p$ & $n$&$r$ &$s$& $\kappa(\bm{\Omega})$&$\beta$&$\rm{Num}_{E}$&$\tfrac{\widehat{E}_p}{n}$&$\tfrac{\widehat{E}_n}{n}$&$\tfrac{E_{\delta}}{n}$\\
		\hline
	Random 1&$1000$ & $999$ & $944$& $10$&$\infty$&$1.3$&$85.14$&$139.13$&$0.12$&$0.005$  \\
		\hline
		Random 1&$1000$ & $999$ & $999$& $2$&$100$&$1$&$63.04$&$4.85$&$0.051$&$0.006$\\
		\hline
		Random 1&$2000$ & $1990$ & $1990$& $11$&$50$&$1$&$43$&$2.04$&$0.07$&$0.001$ \\
		\hline
		Random 1*&$2000$ & $1995$ & $1995$& $10$&$1.94$&$1.05$&$82.5$&$0.08$&$0.05$&$0.002$ \\
		\hline
		Random 2 &$1000$&$995$&$995$&$6$&$483$&$1.2$&$59.3$&$2.72$ &$0.03$&$0.002$\\
		\hline
		High-pass Daubechies wavelet (2 decom. level)&$2048$ & $1024$ & $1023$&$600$& $\infty$&$1.3$&$101.64$&$22.7$&$0.1$&$0.005$ \\
		\hline
	\end{tabular}
\caption{We examine the function $f=\|\bm{\Omega}\cdot\|_1$ when $\bm{\Omega}$ is a tall analysis operator. Three types of operators are considered: Random 1, Random 2 and wavelet. These operators are constructed using the procedure proposed in Items \ref{item.random1}, \ref{item.random2}, and \ref{item.wavelet} in Section \ref{section.simulation}, respectively. For the wavelet case, we construct a $3072\times 1024$ Daubechies wavelet transform where we only retain its high-pass coefficients (of size $2048\times 1$). The number of decomposition levels in the wavelet transformation is two. {\color{\change} Except for Random 1*, we} construct cosparse signals according to the procedure explained in Section \ref{section.simulation}.  {\color{\change}For Random 1* we use $\bm{x}=\bm{P}_{{\rm null}(\bm{\Omega}_{\overline{\mathcal{S}}})} \bm{c}$, where $\bm{c}$ is uniformly distributed on the unit sphere $\mathbb{S}^{n-1}$. In this table, $\rm{Num}_{E}$ denotes the numerator of $\widehat{E}_p$ and is equal to $2\sup_{s\in \partial f(\bm{x})}\|s\|_2$.}
}\label{table.model1}
}
\end{table*}

\begin{table*}[t]

	{\color{\change}
	\centering
	\begin{tabular}{|c|c|c|c|c|c|c|c|c|c|}
		\hline
		Type of Analysis operator $\bm{\Omega}$&$p$ & $n$&$r$&$s$&$\beta$&$\rm{Num}_{E}$&$\tfrac{\widehat{E}_p}{n}$&$\tfrac{\widehat{E}_n}{n}$&$\tfrac{E_{\delta}}{n}$\\ 
		\hline
		Random 1&$1490$ & $1500$ & $1490$& $5$&$1$&$71.76$&$51.93$ &$0.056$&$0.001$\\
		\hline
		Random 1&$1999$&$2000$&$1999$&$10$&$1$&$83.06$&$5.8$&$0.0136$&$0.008$ \\
		\hline
		Random 1 &$1999$&$2000$&$1997$&$10$&$1$&$82.72$&$18.38$&$0.04$&$0.001$\\
		\hline
		Random 2&$1999$&$2000$&$1999$&$10$&$1.05$&$124.3$&$16.15$&$0.008$&$0.002$ \\
		\hline
		TV&$1999$&$2000$ &$1999$ &$350$ &$1$&$124.76$&$9.94$&$0.08$&$0.001$ \\
		\hline
		TV&$1999$&$2000$ &$1999$ &$10$&$1$&$123.72$&$891.3$&$0.003$&$0.001$ \\
		\hline
		TV*&$1999$&$2000$ &$1999$ &$10$&$1$&$123.12$&$0.2$&$0.002$&$0.001$ \\
		\hline
		Low-pass Daubechies wavelet (1 decom. level)&$2048$&$2048$ &$2047$ &$10$ &$1.05$&$85.8$&$130.36$&$0.071$&$0.002$ \\
		\hline
		High-pass Daubechies wavelet (1 decom. level)&$1024$&$1024$ &$1023$ &$8$ &$1.08$&$63.4$&$13.9$&$0.1$&$0.005$ \\
		\hline
	\end{tabular}
\caption{We examine the function $f=\|\bm{\Omega}\cdot\|_1$ when $\bm{\Omega}$ is a fat analysis operators. Four types of  operators are considered: Random 1, Random 2, finite difference, and wavelet. The cases of Random 1 and 2 are constructed using the procedure explained in Items \ref{item.random1} and \ref{item.random2} in Section \ref{section.simulation}, respectively. We build the wavelet matrices with  Daubechies structure where we retain low- and high-pass coefficients. The considered number of decomposition levels is $1$. As it is clear, our error bound $\widehat{E}_n$ outperforms the previous error estimate \eqref{eq.errorbound} denoted by $\widehat{E}_p$ in all cases. 
{\color{\change} Except for TV*,} we construct cosparse signals according to the procedure explained in Section \ref{section.simulation}.
{\color{\change}For TV* we use $\bm{x}=\bm{P}_{{\rm null}(\bm{\Omega}_{\overline{\mathcal{S}}})} \bm{c}$, where $\bm{c}$ is uniformly distributed on the unit sphere $\mathbb{S}^{n-1}$. In this table, $\rm{Num}_{E}$ denotes the numerator of $\widehat{E}_p$ and is equal to $2\sup_{s\in \partial f(\bm{x})}\|s\|_2$.}
}
\label{table.model2}
}
\end{table*}

\subsection{Weak decomposability condition}\label{sec.foygel_fail}
For functions $f=\{\|\cdot\|_1,\|\cdot\|_{1,2},\|\cdot\|_*\}$, one can always find a vector $\bm{z}_0\in\partial f(\bm{x})$ such that the weak decomposability assumption (\ref{eq.weakdecom}) holds. More precisely, one can use \cite[Definition 2]{candes2013simple}:
\begin{align}\label{eq.z0}
&\bm{z}_0={\rm sgn}(\bm{x})~~~\text{for}~ f=\|\cdot\|_1,\nonumber\\
&\bm{z}_0=\frac{\bm{x}_{\mathcal{V}_b}}{\|\bm{x}_{\mathcal{V}_b}\|_2}~~~\text{for}~f=\|\cdot\|_{1,2},\nonumber\\
&\bm{Z}_0=\bm{U}_{n_1\times r}\bm{V}_{n_2\times r}^H~~~\text{for}~f=\|\cdot\|_*,
\end{align}
where $\bm{U}_{n_1\times r}$ and $\bm{V}_{n_2\times r}$ are the bases corresponding to the reduced singular value decomposition of the ground-truth matrix $\bm{X}:=\bm{U}_{n_1\times r}\bm{\Sigma}_{r\times r}\bm{V}_{r\times n_2}^{\rm H}$. {\color{\change}The sets $\{\mathcal{V}_b\}_{b=1}^q$ stand for a partitioning of $\{1,..., n\}$ into $q$ blocks of equal length.} In \cite{zhang2016precise}, it is shown that $f=\|\cdot\|_{\mathrm{TV}}$ satisfies the weak decomposability assumption (\ref{eq.weakdecom}), i.e., there exists $\bm{z}_0\in\partial\|\cdot\|_{\rm TV}(\bm{x})$ which satisfies (\ref{eq.weakdecom}).  
In Section \ref{sec.betacomputations}, we show that the weak decomposability condition does not necessarily hold for the general family of $f=\|\bm{\Omega}\cdot\|_1$; in particular, we construct counter-examples for the case of full-rank tall analysis operators.

\section{Main results}\label{section.mainresult} 
Our main results which are stated in the following theorem, estimate the distance between $\delta(\mathcal{D}(f,\bm{x}))$ and its corresponding upper-bound. 
\begin{thm}\label{thm.maintheorem}
Let $f$ be a proper convex function that promotes the structure of $\bm{x}\neq \bm{0}\in\mathbb{R}^n$ and let $\bm{g}\in\mathbb{R}^n$ be a standard i.i.d Gaussian vector. Suppose $\partial f(x)$ satisfies
\begin{align}\label{eq.mycondition}
&\exists \bm{z}_0~~\mathrm{s.t.}~ \langle \bm{z}-\bm{z}_0,\bm{z}_0 \rangle=0,~~ \forall \bm{z}\in \partial f(\bm{x}). 
\end{align}
Then for any positive values of $\lambda,\zeta$, we have that 
\begin{align}\label{eq.main}
0\le E_{\delta} \le &(4\lambda\beta+\gamma)~ \omega(\mathcal{D}(f,\bm{x})\cap \mathbb{B}^n)+\gamma(\zeta+2\lambda\beta)+4\lambda^2\beta^2\nonumber\\
&:=\widehat{E}_n,
\end{align}
and
\begin{align}\label{eq.generalizedfoygel}
0\le\inf_{t\ge0}\mathds{E}~\mathrm{dist}(\bm{g},t\partial f(\bm{x}))- \omega(\mathcal{D}(f,\bm{x})\cap \mathbb{B}^n)\le 1.6+4\beta,
\end{align}
where $\gamma$ is the constant
\begin{align}\label{eq.gamma}
\gamma=\sqrt{72}\sqrt{\ln\frac{3}{1-4e^{-\frac{\lambda^2}{2}}-2e^{-\frac{\zeta^2}{2}}}},
\end{align}
and $\beta$ is given by
\begin{align}\label{eq.beta}
\beta=\frac{\|\bm{z}_1\|_2}{\|\bm{z}_0\|_2},
\end{align}
where
\begin{align}\label{eq.z1}
\bm{z}_1=\arg\min_{\bm{z}\in\partial f(\bm{x})}\|\bm{z}\|_2.
\end{align}
\end{thm}
Proof. See Appendices \ref{maintheoremproof} and \ref{proof.genelarizedfoygel}.

{\color{\change}\begin{rem}\label{rem.2}
The error estimate in \eqref{eq.main} is the main result of Theorem \ref{thm.maintheorem}, while \eqref{eq.generalizedfoygel} can be thought of as an extension of Result \ref{prop.foygel} to more general structure-inducing functions (including $\ell_1$-analysis). Despite the similarities between  $\inf_{t\ge0}\mathds{E}~\mathrm{dist}(\bm{g},t\partial f(\bm{x}))$ and $\inf_{t\ge0} \sqrt{\mathds{E}~\mathrm{dist}^2(\bm{g},t\partial f(\bm{x}))}$, these two terms are different, and \eqref{eq.generalizedfoygel} cannot be considered as an error bound. We should add that \eqref{eq.generalizedfoygel} is used in our proof of \eqref{eq.main}.
%
\end{rem}}
\begin{rem}
If $\beta$ is bounded, as $\omega(\mathcal{D}(f,\bm{x})\cap \mathbb{B}^n)\le\sqrt{n}$, $\tfrac{\widehat{E}_n}{n}$ asymptotically tends to $0$. We numerically observe in Section \ref{sec.betacomputations} that for functions $f=\{\|\cdot\|_1,\|\cdot\|_{1,2},\|\cdot\|_*,\|\bm{\Omega}\cdot\|_1\}$, $\beta$ is bounded in most cases of $\bm{\Omega}$. In case of $\ell_1$ analysis, $\beta$ can be upper-bounded by a function of the \textit{generalized sign vector} of $\bm{x}$ (i.e. $\bm{\Omega}^T{\rm sgn}(\bm{\Omega x})$).
\end{rem}

{\color{\change}When $\beta$ is bounded in the $\ell_1$-analysis case, our bound in Theorem \ref{thm.maintheorem} implies that the normalized error gap is vanishing asymptotically; equivalently, it implies that $U_{\delta}$ is a good estimate of $\delta\big(\mathcal{D}(\|\bm{\Omega}\cdot\|_1,\bm{x})\big)$. Our result holds for various analysis operators including the ones that have non-trivial linear dependencies among their rows. It should be noted that we do not guarantee the boundedness of $\beta$; in case $\beta$ fails to remain bounded, our error estimate is no longer effective.
%
}


\subsection{Evaluation of $\beta$}\label{sec.betacomputations}
To compute $\beta$, we need both $\bm{z}_1$ in (\ref{eq.z1}) and $\bm{z}_0$ in (\ref{eq.mycondition}). It is not difficult to see that for functions $f=\{\|\cdot\|_1,\|\cdot\|_{1,2},\|\cdot\|_*\}$, $\bm{z}_1$ in \eqref{eq.z1} is obtained by
\begin{align}\label{eq.z1foreasyfunctions}
&\bm{z}_1={\rm sgn}(\bm{x})~~~\text{for}~ f=\|\cdot\|_1,\nonumber\\
&\bm{z}_1=\frac{\bm{x}_{\mathcal{V}_b}}{\|\bm{x}_{\mathcal{V}_b}\|_2}~~~\text{for}~f=\|\cdot\|_{1,2},\nonumber\\
&\bm{Z}_1=\bm{U}_{n_1\times r}\bm{V}_{n_2\times r}^H~~~\text{for}~f=\|\cdot\|_*,
\end{align}
where $\bm{U}_{n_1\times r}$ and $\bm{V}_{n_2\times r}$ are bases corresponding to the reduced singular value decomposition of the ground-truth matrix $\bm{X}:=\bm{U}_{n_1\times r}\bm{\Sigma}_{r\times r}\bm{V}_{r\times n_2}^{\rm H}$. 

Choosing $\bm{z}_0=\bm{z}_1$ in the subdifferential is common for functions $f=\{\|\cdot\|_1,\|\cdot\|_{1,2},\|\cdot\|_*\}$\cite{candes2013simple,foygel2014corrupted}.
Unlike these simple choices, obtaining $\bm{z}_0$ for the cosparse vectors is more involved. In the following proposition, we discuss this issue when $f=\|\bm{\Omega}\cdot\|_1$, where $\bm{\Omega}$ is either a tall or a fat analysis operator. 
 
\begin{prop}\label{prop.l1analysis}
	Consider the cosparse vector $\bm{x}\in\mathbb{R}^n$ in the analysis domain $\bm{\Omega}\in\mathbb{R}^{p\times n}$  with support $\mathcal{S}$. Then, 
	\begin{align}\label{eq.z0prop}
&\bm{z}_0=\bm{P}_{{\rm null}(\bm{\Omega}_{\overline{\mathcal{S}}})}\bm{\Omega}^T{\rm sgn}(\bm{\Omega} \bm{x}),
	\end{align}
satisfies (\ref{eq.mycondition}).
\end{prop}
Proof. See Appendix \ref{proof.propl1ana}.

In what follows, we examine some special and important implications of this proposition.
\begin{rem}
In case of $\bm{\Omega}=\bm{I}$ (i.e. $f=\|\cdot\|_1$), $\bm{z}_0={\rm sgn}(\bm{x})$. This supports the fact that choosing $\bm{z}_0$ in the subdifferential is reasonable and efficient. 
\end{rem}

\begin{rem}(Upper-bound on $\beta$)
Employing \eqref{eq.z0prop}, we can express $\beta$ as
\begin{align}\label{eq.upper_on_beta}
\beta &:=\frac{\inf_{\|\widetilde{\bm{z}}_{\overline{\mathcal{S}}}\|_{\infty}\le 1}\|\bm{\Omega}^{\rm T}{\rm sgn}(\bm{\Omega} \bm{x})+\bm{\Omega}_{\overline{\mathcal{S}}}^{\rm H}\widetilde{\bm{z}}_{\overline{\mathcal{S}}}\|_2}{\|\bm{P}_{{\rm null}(\bm{\Omega}_{\overline{\mathcal{S}}})}\bm{\Omega}^T{\rm sgn}(\bm{\Omega}\bm{x})\|_2}\nonumber\\
&\le \frac{\|\bm{\Omega}_{\mathcal{S}}^T{\rm sgn}(\bm{\Omega}_{\mathcal{S}}\bm{x})\|_2}{\|\bm{P}_{{\rm null}(\bm{\Omega}_{\overline{\mathcal{S}}})}\bm{\Omega}_{\mathcal{S}}^T{\rm sgn}(\bm{\Omega}_{\mathcal{S}}\bm{x})\|_2}.
\end{align}	
The boundedness of $\beta$ is consistently observed in our numerical results (see Tables \ref{table.model1} and \ref{table.model2}). We also prove the boundedness for some special cases of $\bm{\Omega}\in\mathbb{R}^{p\times n}$. For instance, for fat matrices with orthogonal rows (a special case of which is investigated in Item \ref{item.random1} of Section \ref{section.simulation}), $\beta$ equals $1$. (see Appendix \ref{proof.beta_one} for the proof). Another example is when the elements of $\bm{\Omega}$ are drawn from an i.i.d. Gaussian distribution. In this case, under the assumption  $\frac{{\color{\change}|\overline{\mathcal{S}}|+1}}{n} \leq \rho <1$, we have that $$\beta\le \frac{1}{\sqrt{1-\rho}}$$ with high probability in high dimensions (see Appendix \ref{beta_gaussian} for the proof).
\end{rem}

\section{Numerical experiments}\label{section.simulation}
\begin{figure}[t]
	\hspace*{-0cm}
	\centering
	\includegraphics[scale=.25]{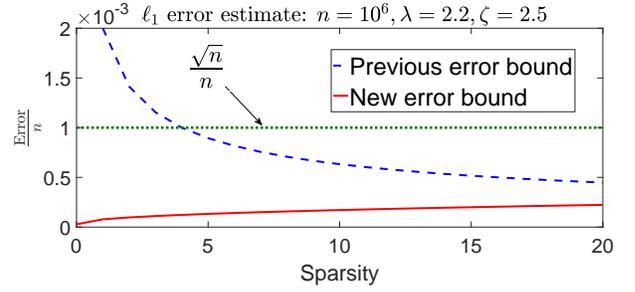}
	\caption{Two strategies of obtaining the error of $\delta(\mathcal{D}(f,\bm{x}))$ from (\ref{eq.upperbound}) in case of $f=\|\cdot\|_1$. The previous and new error bounds come from (\ref{eq.ell1}) and (\ref{eq.main}), respectively. }
	\label{fig.l1}
\end{figure}

\begin{figure}[t]
	\hspace*{-.4cm}
	\centering
	\includegraphics[scale=.25]{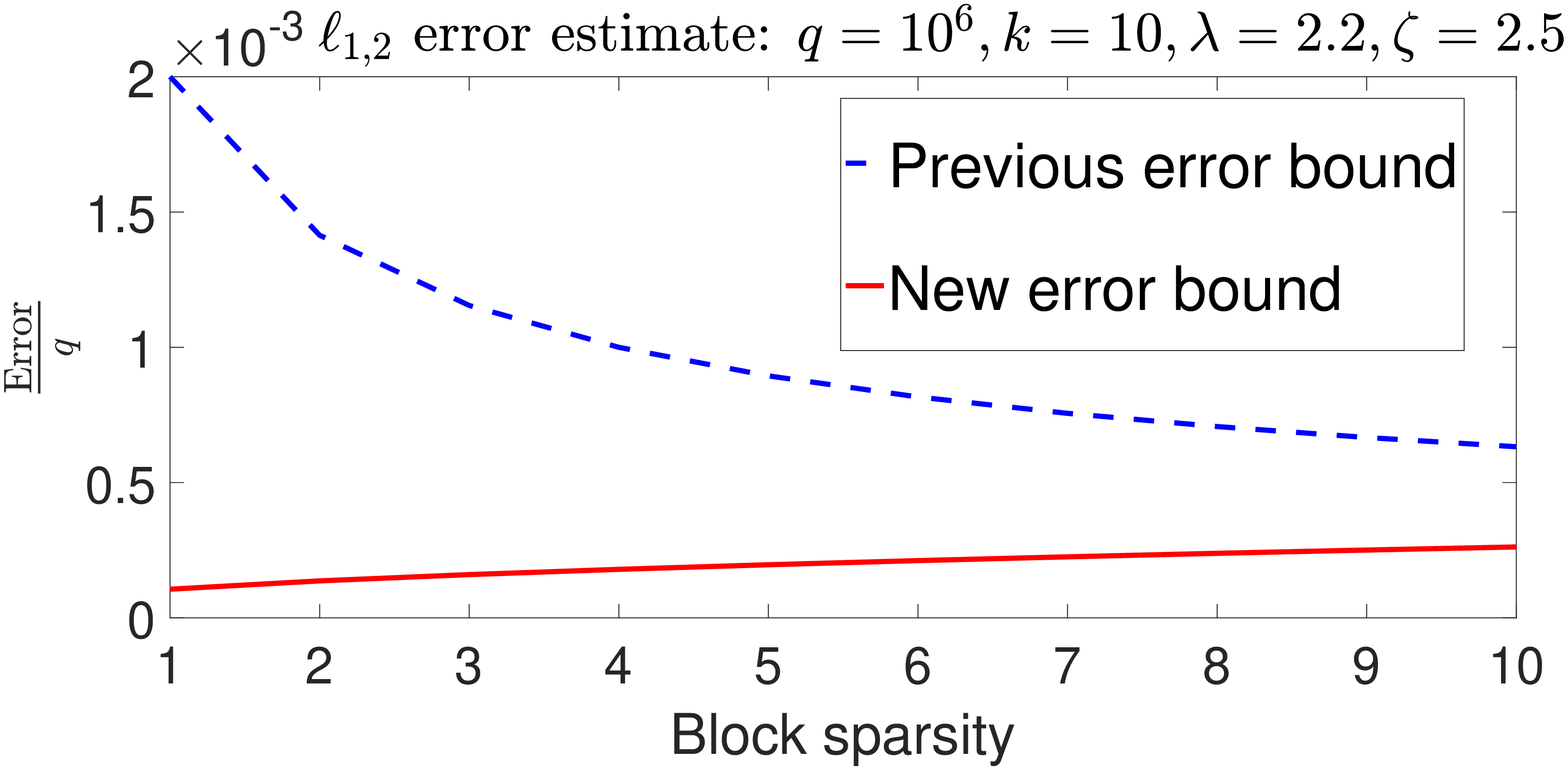}
	\caption{Comparison of the error (\ref{eq.error_delta}) in case of $f=\|\cdot\|_{1,2}$. The previous and new error bounds come from (\ref{eq.ell12}) and (\ref{eq.main}), respectively. }
	\label{fig.l12}
\end{figure}

\begin{figure}[t]
	\hspace*{-.4cm}
	\centering
	\includegraphics[scale=.28]{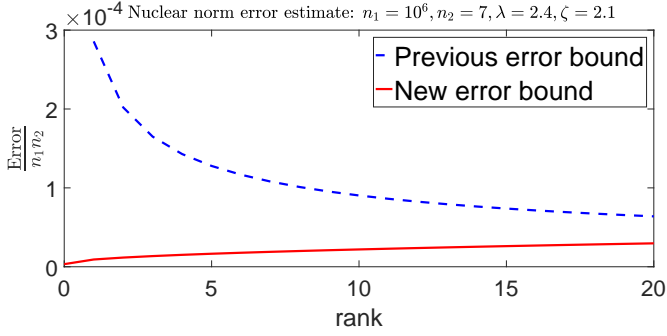}
	\caption{Comparison of the error (\ref{eq.error_delta}) in case of $f=\|\cdot\|_{*}$. The previous and new error bounds come from \eqref{eq.nuclear2} and (\ref{eq.main}), respectively.}
	\label{fig.nuc}
\end{figure}

In this section, we numerically compare the new error bound of (\ref{eq.main}) against the bound (\ref{eq.errorbound}) derived using the existing approach for various low-dimensional structures. For each test, we optimize $\lambda$ and $\zeta$ to minimize the right-hand side of (\ref{eq.main}). Figures \ref{fig.l1}, \ref{fig.l12}, and \ref{fig.nuc} show the {\color{\change} proposed error bound \eqref{eq.main} and the error estimates \eqref{eq.ell1}, \eqref{eq.ell12}, and \eqref{eq.nuclear2}, for $\|\cdot\|_1$, $\|\cdot\|_{1,2}$ and $\|\cdot\|_{*}$, respectively.} In all cases, the sparsity/rank values are set very small. To compute $\omega(\mathcal{D}(f,\bm{x})\cap \mathbb{B}^n)$ in (\ref{eq.main}), we used its upper-bound obtained via (\ref{eq.lazem}), \cite[Equations D.6, D.10]{amelunxen2013living}, and \cite[Lemma 1]{daei2018exploiting}. It is clear from these figures that the new error bound outperforms the previous error bound (\ref{eq.errorbound}) in very low sparsity/rank regimes;
it should be emphasized that the curves depict the upper-bound of (\ref{eq.main}). {\color{\change} Notice that in these three cases, both $\frac{\widehat{E}_p}{n}$ and $\frac{E_n}{n}$ tend to zero at large $n$.}

Due to the varying nature of the $\ell_1$ analysis case, we construct three kinds of analysis operators as follows:
	\begin{enumerate}
		\item \underline{Random 1}: We first generate a $p\times n$ Gaussian matrix with i.i.d. elements. Then, we compute its SVD as $\bm{U}_{p\times p}\bm{\Sigma}_{p\times n}\bm{V}_{n\times n}^H$. Then, $\bm{\Sigma}$ is replaced with the matrix 
		\begin{align}
		\bm{\Sigma}_1:=\begin{bmatrix}
		\bm{I}_r&\bm{0}_{r\times n-r}\\
		\bm{0}_{p-r\times r}&\bm{0}_{p-r\times n-r}
		\end{bmatrix}
		\end{align}
		to get
		\begin{align}\label{eq.random1_cons}
		\bm{\Omega}=\bm{U}_{p\times p}\bm{\Sigma}_1\bm{V}_{n\times n}^{\rm H}.
		\end{align}
		{\color{\change}When $r=n\leq p$, the constructed $\bm{\Omega}$ in \eqref{eq.random1_cons} is a tight frame. To have an analysis operator with more varied singular values, we proceed with
		\begin{align}\label{eq.random1operators}
			\bm{\Omega}=\bm{D}_{p\times p}\bm{U}_{p\times p}\bm{\Sigma}_1\bm{V}_{n\times n}^{\rm H},
		\end{align}
		where $\bm{D}_{p\times p}$ is a diagonal matrix.}
		This type of matrices is widely used as a benchmark in \cite{kabanava2015analysis,nam2013cosparse,genzel2017ell}. The above approach was directly adopted from \cite{nam2013cosparse}.
\label{item.random1}
		
		
		\item \underline{Random 2}: In this case, we simply use the Gaussian ensemble by constructing a $p\times n$ matrix with i.i.d. elements {\color{\change}that each follows $\mathcal{N}(0,\sigma^2)$}.
		Here, $\sigma^2$ implicitly specifies the range of the singular values.\label{item.random2}
		
		\item \underline{Wavelet}: We choose a redundant wavelet transform from the package SPOT \cite{spot} to construct an analysis operator. The wavelet filter is chosen from the Daubechies family and has length $8$. In some cases, we retain only the high-pass or low-pass coefficients. The rows of the wavelet operator $\bm{\Omega}$ might have non-trivial linear dependencies. \label{item.wavelet}
	\end{enumerate}
	
	 {\color{\change}For a general analysis operator $\bm{\Omega}_{p\times n}$, we first randomly select $\mathcal{S}$ among the subsets of $\{1,...,p\}$ with size $s$. Then, we check whether $\bm{\Omega}_{\mathcal{S}}\bm{P}_{{\rm null}(\bm{\Omega}_{\overline{\mathcal{S}}})}$ is the trivial $\bm{0}$ operator or not. In the trivial case, we regenerate $\mathcal{S}$ and repeat the test; otherwise, we form $\bm{x}$ via
	$$\bm{x}=\bm{P}_{{\rm null}(\bm{\Omega}_{\overline{\mathcal{S}}})} \bm{w},$$
	where $\bm{w}$ is such that $\bm{\Omega} \bm{x} \neq \bm{0}$.
	}
In this paper, as we would like to highlight the difference between the new error bound and the existing ones, we focus on a subclass of analysis sparse vectors. More specifically, whenever there are non-trivial linear dependencies among the columns of the analysis operator (fat or tall), we generate $\bm{x}$ according to the procedure explained in Section \ref{sec.ame_fail}. Whenever the analysis operator	is of type Random 2 (with highly coherent rows), we generate $\bm{x}$ as
		 \begin{align}\label{eq.random2_sig_gen}
		 \bm{x}=\bm{P}_{{\rm null}({\bm{\Omega}}_{\overline{\mathcal{S}}})}(\bm{w}+\alpha \bm{v}_r),
		 \end{align}
		 where $\bm{w}\in\mathbb{R}^n$ and $\alpha\in\mathbb{R}$ are arbitrary quantities and
	%
%
%
$\bm{v}_r$ is the right singular vector of ${\bm{\Omega}}_{\mathcal{S}}\bm{P}_{{\rm null}({\bm{\Omega}}_{\overline{\mathcal{S}}})}$ corresponding to the minimum non-zero singular value ({\color{\change}indeed, $r={\rm rank}({\bm{\Omega}}_{\mathcal{S}}\bm{P}_{{\rm null}({\bm{\Omega}}_{\overline{\mathcal{S}}})})$}). 
With this choice, the denominator of the error bound \eqref{eq.errorbound} simplifies to
		\begin{align}\label{eq.analysis.model1}
		\frac{\|{\bm{\Omega}}_{\mathcal{S}}\bm{P}_{{\rm null}({\bm{\Omega}}_{\overline{\mathcal{S}}})}\bm{w}+\alpha\sigma_r\bm{u}_r\|_1}{\|\bm{P}_{{\rm null}({\bm{\Omega}}_{\overline{\mathcal{S}}})}\bm{w}+\alpha \bm{P}_{{\rm null}({\bm{\Omega}}_{\overline{\mathcal{S}}})} \bm{v}_r\|_2},
		\end{align}
		where $\bm{u}_r$ is the $r$th left singular vector of  ${\bm{\Omega}}_{\mathcal{S}}\bm{P}_{{\rm null}({\bm{\Omega}}_{\overline{\mathcal{S}}})}$. For highly coherent analysis operators where the minimum singular value is very small, by increasing $\alpha$, the denominator of the error bound \eqref{eq.errorbound} is likely to decrease. 
		
{\color{\change}For the tall analysis operators of type Random 1 in \eqref{eq.random1operators}, we construct pairs of $(\bm{\Omega},\bm{x})$ as follows. We set
\begin{align}
&\bm{\Omega}_{\mathcal{S}}=\bm{D}_{1} \, \bm{U}_{|\mathcal{S}|\times n}\bm{V}^{\rm H},\nonumber\\
&\bm{\Omega}_{\overline{\mathcal{S}}}=\bm{D}_{2} \, \bm{U}_{|\overline{\mathcal{S}}|\times n}\bm{V}^{\rm H},
\end{align}
where $\bm{D}_{1}$ and $\bm{D}_{2}$ are arbitrary diagonal matrices with non-negative values of size $|\mathcal{S}|\times |\mathcal{S}|$ and $|\overline{\mathcal{S}}|\times |\overline{\mathcal{S}}|$, respectively,
%
 and $\bm{U}_{|\mathcal{S}|\times n}$ is a sub-matrix of $\bm{U}_{p\times p}$ restricted to the rows and columns in $\mathcal{S}$ and $[n]$, respectively. We further generate $\bm{x}$  as
\begin{align}
\bm{x}=\bm{P}_{{\rm null}(\bm{U}_{|\overline{\mathcal{S}}|\times n}\bm{V}^{\rm H})} (\bm{w}+\alpha \bm{v}_{\min}^{\prime}),
\end{align}
where $\alpha>0$ is an arbitrary real, $\bm{w}$ is an arbitrary vector in $\mathbb{R}^n$ and $\bm{v}_{\min}^{\prime}$ is the right singular vector corresponding to the minimum singular value $\sigma_{\min}$ of $\bm{U}_{|\mathcal{S}|\times n}\bm{V}^{\rm H}\bm{P}_{{\rm null}(\bm{U}_{|\overline{\mathcal{S}}|\times n}\bm{V}^{\rm H})}$. Then, the denominator of the error bound \eqref{eq.errorbound} becomes
\begin{align}
&\frac{\|\bm{\Omega}_{\mathcal{S}}\bm{x}\|_1}{\|\bm{x}\|_2}=\nonumber\\
&\tfrac{\|\bm{D}_{1}\bm{U}_{|\mathcal{S}|\times n}\bm{V}^{\rm H}\bm{P}_{{\rm null}(\bm{U}_{|\overline{\mathcal{S}}|\times n}\bm{V}^{\rm H})} \bm{w}+\alpha \sigma_{\min}\bm{D}_{1} \bm{u}_{\min}^{\prime}\|_1}{\|\bm{P}_{{\rm null}(\bm{U}_{|\overline{\mathcal{S}}|\times n}\bm{V}^{\rm H})} \bm{w}+\alpha \bm{P}_{{\rm null}(\bm{U}_{|\overline{\mathcal{S}}|\times n}\bm{V}^{\rm H})}\bm{v}_{\min}^{\prime}\|_2},
\end{align}
where $\bm{u}_{\min}^{\prime}$ is the left singular vector corresponding to $\sigma_{\min}$. 
As we have full control over $\alpha$ and the diagonal elements of $\bm{D}_1$, we can make the error bound  \eqref{eq.errorbound}  arbitrarily large (decreasing the diagonal elements of $\bm{D}_1$ while increasing $\alpha$).
}

Table \ref{table.model1} compares the two error bounds for various examples of tall analysis operators. We observe that our error bound \eqref{eq.main} is considerably superior to the error estimate \eqref{eq.errorbound}. We use three kinds of analysis operators: Random 1, 2 and Daubechies wavelet for different sparsity levels and dimensions. Notice that the Daubechies wavelet of size $2048\times 1024$ is constructed by retaining the high-pass components of a $2$-level Daubechies wavelet of size $3072\times 1024$. The wavelet transformation is computed by the SPOT package \cite{spot}. The procedure of computing $\omega(\mathcal{D}(\|\bm{\Omega}\cdot\|_{1},\bm{x})\cap \mathbb{B}^n)$ in \eqref{eq.main} is explained in Appendix \ref{sec.TrueGaussianwidth}.

In Table \ref{table.model2}, we examine fat analysis operators including Random 1 and 2 structures, TV and Daubechies wavelet. Again, our bound \eqref{eq.main} confirms that $U_{\delta}$ is close to $\delta(\mathcal{D}(\|\bm{\Omega} \cdot\|_1,\bm{x}))$, while the error bound \eqref{eq.errorbound} is inconclusive.

{\color{\change}Different from the above mentioned strategies for generating analysis-sparse signals, we also construct signals (shown by Random 1* in Table \ref{table.model1} and TV* in Table \ref{table.model2}) as $\bm{x}=\bm{P}_{{\rm null}(\bm{\Omega}_{\overline{\mathcal{S}}})}\bm{c}$ where $\bm{c}$ is uniformly distributed on the unit sphere. In these cases, we observe that the error estimate \eqref{eq.errorbound} is effective.}

\section{Conclusion}\label{section.conclusion}
In this work, we presented an error estimate bound for the statistical dimension. This new bound shows that the statistical dimension is well described by its common upper-bound \eqref{eq.upperbound} in {\color{\change} some settings of}   TV structure and $\ell_1$ analysis.

\appendices

\section{Proof of Theorem \ref{thm.maintheorem} (\ref{eq.main})}\label{maintheoremproof}
{\color{\change}Before beginning the proof, we define some parameters and provides a proof sketch to enhance the readability. Given $\lambda>0$, we define the parameters
\begin{align}
&\alpha:=\mathds{E}[t_{\bm{g}}]+\frac{\lambda}{\|\bm{z}_0\|_2},\label{eq.alpha}\\
&t_{\bm{g}}:=\arg\min_{t\ge 0}\mathrm{dist}(\bm{g},t\partial f(\bm{x})),\label{eq.tg2}
\end{align}
and the function
\begin{align}
\label{eq.phig}
&\phi(\bm{g}):=\mathrm{dist}^2(\bm{g},\alpha \partial f(\bm{x}))-\mathrm{dist}^2(\bm{g},\mathrm{cone}(\partial f(\bm{x}))).
\end{align}
Notice that due to \cite[Lemma C.1]{amelunxen2013living}, whenever $f$ is a proper convex function (as is the case in this paper), $t_{\bm{g}}$ is well-defined and unique.
Define the event
\begin{align}\label{eq.event}
\mathcal{E}=\Big\{|t_{\bm{g}}-\mathds{E}[t_{\bm{g}}]|<\frac{\lambda}{\|\bm{z}_0\|_2}\Big\}.
\end{align}
For fixed $\bm{g}$, by considering the condition \eqref{eq.mycondition},  $t_{\bm{g}}$ is a $\frac{1}{\|\bm{z}_0\|_2}$ Lipschitz function of $\bm{g}$ \cite[Lemma 3]{foygel2014corrupted}\footnote{The proof of \cite[Lemma 3]{foygel2014corrupted} does not need $\bm{z}_0$ to be an element of $\partial f(\bm{x})$.}. Hence, by a concentration inequality for Lipschitz functions of Gaussian vectors (see, for example, \cite[Theorem 8.40]{foucart2013mathematical}), we get that
\begin{align}\label{eq.concen1}
\mathds{P}\{\mathcal{E}\}\ge p_0:=1-2e^{-\frac{\lambda^2}{2}}.
\end{align}
\begin{sketch}
	The goal is to find an upper-bound for $E_{\delta}$.
	Instead of bounding $E_{\delta}$, we bound the expression
	\begin{align}
	&\mathds{E}~\mathrm{dist}^2(\bm{g},\alpha\partial f(\bm{x}))-\delta(\mathcal{D}(f,\bm{x}))=\nonumber\\
	&\mathds{E}~\mathrm{dist}^2(\bm{g},\alpha\partial f(\bm{x}))-\mathds{E}~\mathrm{dist}^2(\bm{g},{\rm cone}(\partial f(\bm{x})))=\mathds{E}\phi(\bm{g})\nonumber,
	\end{align}
	which is counted as its upper-bound.
	To reach this goal, we do the following steps:
	\begin{enumerate}
		
		\item When $\mathcal{E}$ holds, we find that $$\phi_1(\bm{g}):=\phi(\bm{g})- 4\lambda \beta \mathrm{dist}(\bm{g},\mathrm{cone}(\partial f(\bm{x})))\le 4\lambda^2\beta^2.$$
		\item \label{it.2} We obtain a lower-bound for the probability of the event $\phi_1(\bm{g})\le 4\lambda^2 \beta^2$.
		
		\item \label{it.3}Then, we obtain a concentration inequality for the expression $\phi_1(\bm{g})$ which is associated with $\mathds{E}\phi_1(\bm{g})$ (see Lemma \ref{lemma.lip}).
		\item Combining the concentration inequality in Item \ref{it.3} and the lower-bound in Item \ref{it.2}, we reach a contradiction unless we have that $\mathds{E}\phi_1(\bm{g})$ is bounded above by a certain expression. 
		\item Finally, the upper-bound on $\mathds{E}\phi(\bm{g})$ (and thus bound on $E_{\delta}$) is directly obtained using the upper-bound on $\mathds{E}\phi_1(\bm{g})$.
	\end{enumerate}
\end{sketch}

We now prove each of the above mentioned parts in details. 
Suppose that $\mathcal{E}$ holds. 
Define $\bm{z}^*$ such that
	\begin{align}
	\mathrm{dist}^2(\bm{g},t_{\bm{g}}\partial f(\bm{x}))=\|\bm{g}-t_{\bm{g}}\bm{z}^*\|_2^2.
	\end{align}
	Take
	\begin{align}\label{eq.zinsub}
	\bm{z}=\frac{t_{\bm{g}}}{\alpha} \bm{z}^*+(1-\frac{t_{\bm{g}}}{\alpha})\bm{z}_1~\in\partial f(\bm{x}),
	\end{align}
	where $\bm{z}_1$ is defined in~\eqref{eq.z1}. 
	That this is an element of $\partial f(\bm{x})$ follows from the fact that both $\bm{z}_1$ and $\bm{z}^*$ are in $\partial f(\bm{x})$, and that $\mathcal{E}$ implies that $t_{\bm{g}}/\alpha<1$.
	Then, we can find an upper-bound for $\phi(\bm{g})$ as follows: 
	\begin{align}\label{eq.upp1}
	&\mathrm{dist}^2(\bm{g},\alpha \partial f(\bm{x}))\le\|\bm{g}-\alpha\bm{z}\|_2^2=\|\bm{g}-t_{\bm{g}}\bm{z}^*+t_{\bm{g}}\bm{z}^*-\alpha\bm{z}\|_2^2\nonumber \\
	&=\|\bm{g}-t_{\bm{g}}\bm{z}^*\|_2^2+\|t_{\bm{g}}\bm{z}^*-\alpha \bm{z}\|_2^2+
	2\langle \bm{g}-t_{\bm{g}}\bm{z}^*, t_{\bm{g}}\bm{z}^*-\alpha \bm{z}\rangle\nonumber \\
	&\le \mathrm{dist}^2(\bm{g},\mathrm{cone}(\partial f(\bm{x})))
	\nonumber \\
	&\quad +
	(t_{\bm{g}}-\alpha)^2\|\bm{z}_1\|_2^2
	+ 2|t_{\bm{g}}-\alpha|\langle \bm{g}-t_{\bm{g}}\bm{z}^*, \bm{z}_1 \rangle \nonumber\\
	&\le \mathrm{dist}^2(\bm{g},\mathrm{cone}(\partial f(\bm{x})))\nonumber \\ 
	&\quad +4\lambda^2\beta^2 +4\lambda\beta\mathrm{dist}(\bm{g},\mathrm{cone}(\partial f(\bm{x}))),
	\end{align}
	where for the last inequality we use the fact that $\mathcal{E}$ holds, the definition of $\beta$~\eqref{eq.beta}, 
	and the Cauchy-Schwartz inequality. In the following, we obtain a lower-bound for the probability of the event $\phi_1(\bm{g})\le 4\lambda^2\beta^2$.
	\begin{align}\label{eq.geomet}
	&\mathds{P}\Big\{\phi_1(\bm{g})\le 4\lambda^2\beta^2\Big\}=\mathds{P}\Big\{\phi_1(\bm{g})\le 4\lambda^2\beta^2\big|\mathcal{E}\Big\}\mathds{P}\{\mathcal{E}\}+\nonumber\\
	&\mathds{P}\Big\{\phi_1(\bm{g})\le 4\lambda^2\beta^2\big|\overline{\mathcal{E}}\Big\}\mathds{P}\{\overline{\mathcal{E}}\}\ge 
	\mathds{P}\{\mathcal{E}\}\ge p_0,
	\end{align}
	where we used \eqref{eq.upp1}, which implies $\mathds{P}\Big\{\phi_1(\bm{g})\le 4\lambda^2\beta^2\big|\mathcal{E}\Big\}=1$, and \eqref{eq.concen1}.
	
	In what follows, we propose a lemma that provides a relation between $\phi_1(\bm{g})$ and $\mathds{E}\phi_1(\bm{g})$.
	\begin{lem}\label{lemma.lip}
		Let $\bm{g}\in\mathbb
		R^n$ be a standard normal i.i.d. vector. Then, for given  $\lambda,\zeta>0$,
		\begin{align}\label{eq.reform}
		\mathds{P}\{\phi_1(\bm{g})-\mathds{E}\phi_1(\bm{g})\le -\gamma(\zeta+\mathds{E}~\mathrm{dist}(\bm{g},\mathrm{cone}\partial f(\bm{x}))\nonumber \\
		\quad +2\lambda \beta)\}\le p_0,
		\end{align}
		where
	\begin{equation*}
	\gamma:=\sqrt{72}\sqrt{\ln\frac{3}{1-4e^{-\frac{\lambda^2}{2}}-2e^{-\frac{\zeta^2}{2}}}},
	\end{equation*}
	as defined in~\eqref{eq.gamma}. 
	\end{lem}
	The proof of this Lemma is postponed to Appendix \ref{proof.lemmalip}.

By considering (\ref{eq.geomet}) and (\ref{eq.reform}), we reach a contradiction unless
	\begin{align}
	\mathds{E}[\phi_1(\bm{g})]\le \gamma (\zeta+2\lambda\beta+\mathds{E}{\rm dist}(\bm{g},{\rm cone} (\partial f(\bm{x}))))+4\lambda^2\beta^2.
	\end{align}
	By expressing $\phi_1$ in terms of $\phi$ and identifying the expected distance to the subdifferential cone as the Gaussian width, we reach the right-hand side of \eqref{eq.main}. The left-hand side is obtained by applying the Jensen's inequality on the infimum of an affine function (which is always concave).
	}

\section{Proof of Theorem \ref{thm.maintheorem} (\ref{eq.generalizedfoygel})}\label{proof.genelarizedfoygel}
{\color{\change}Our approach in this part is to a great extent, similar to the proof of Result \ref{prop.foygel}. However, the difference lies in the fact that \cite[Proposition 1]{foygel2014corrupted} needs the condition \eqref{eq.weakdecom}, while our proof needs the condition \eqref{eq.mycondition} which holds for more general structure-inducing functions including $\ell_1$ analysis. However, this bound is not an error estimate for the task of predicting the phase transition (see the explanations in Remark \ref{rem.2}) and is used in our analysis in proving \eqref{eq.main}. We proceed with an overview of the proof.
We first define 
\begin{align}\label{eq.phi1g}
\phi_2(\bm{g}):=\mathrm{dist}(\bm{g},\alpha \partial f(\bm{x}))-\mathrm{dist}(\bm{g},\mathrm{cone}\partial f(\bm{x})).
\end{align}
\begin{sketch}
	The goal is to find an upper-bound for the expression
	\begin{align}
	\inf_{t\ge0}\mathds{E}~\mathrm{dist}(\bm{g},t\partial f(\bm{x}))- \omega(\mathcal{D}(f,\bm{x})\cap \mathbb{B}^n).
	\end{align}
We instead intend to find an upper-bound for $\mathds{E}\phi_2(\bm{g})$. To reach this goal, we follow the below steps:
\begin{enumerate}
	
	\item Under the assumption that the event $\mathcal{E}$ holds, we find that $\phi_2(\bm{g})\le 2\lambda \beta$.
	\item \label{it2.2} We obtain a lower-bound for the probability of the event $\phi_2(\bm{g})\le 2\lambda \beta$.
	
	\item \label{it2.3}We obtain a concentration inequality for the expression $\phi_2(\bm{g})$ which is a $2$-Lipschitz function of $\bm{g}$.
	\item The concentration inequality in Item \ref{it2.3} and the lower-bound in Item \ref{it2.2} contradict each other unless we have that $\mathds{E}\phi_2(\bm{g})$ is bounded above by a certain expression. 

\end{enumerate}
\end{sketch}

We now provide the details of the proof. Suppose that \eqref{eq.event} holds. Define $\bm{z}^*$ such that
\begin{align}
\mathrm{dist}(\bm{g},t_{\bm{g}}\partial f(\bm{x}))=\|\bm{g}-t_{\bm{g}}\bm{z}^*\|_2.
\end{align}
By recalling  \eqref{eq.zinsub} and  \eqref{eq.alpha}, we have that
\begin{align}\label{eq.generalized}
&\mathrm{dist}(\bm{g},\alpha \partial f(\bm{x}))\le\|\bm{g}-\alpha\bm{z}\|_2=\|\bm{g}-t_{\bm{g}}\bm{z}^*+t_{\bm{g}}\bm{z}^*-\alpha\bm{z}\|_2\nonumber\\
&\le \|\bm{g}-t_{\bm{g}}\bm{z}^*\|_2+\|t_{\bm{g}}\bm{z}^*-\alpha\bm{z}\|_2\le \|\bm{g}-t_{\bm{g}}\bm{z}^*\|_2+\nonumber\\
&|t_{\bm{g}}-\alpha|\|\bm{z}_1\|_2\le \mathrm{dist}(\bm{g},\mathrm{cone}\partial f(\bm{x}))+2\lambda\beta.
\end{align}
Since $\mathds{P}\{\mathcal{E}\}\ge p_0$, we have that 
\begin{align}\label{eq.geomet1}
\mathds{P}\{\phi_2(\bm{g})\le 2\lambda \beta \}\ge p_0,
\end{align}
by the argument in \eqref{eq.geomet}. Moreover, since $\phi_2(\bm{g})$ is a $2$-Lipschitz function of $\bm{g}$,  the concentration inequality for Lipschitz functions \cite[Theorem 8.40]{foucart2013mathematical}, implies that
\begin{align}
&\mathds{P}\Big{\{}\phi_2(\bm{g})-\mathds{E}[\phi_2(\bm{g})]\le -r\Big{\}}\le e^{-\frac{r^2}{8}}.
\end{align}
With a change of variables, we reach:
\begin{align}\label{eq.lipmamoli}
&\mathds{P}\Big\{\phi_2(\bm{g})-\mathds{E}[\phi_2(\bm{g})]\le -\sqrt{8\ln \frac{1}{p_0}}\Big\}\le p_0.
\end{align}
Note that \eqref{eq.geomet1} and \eqref{eq.lipmamoli} contradict each other unless
\begin{align}
&\mathds{E}\{\phi_2(\bm{g})\}\le \sqrt{8\ln \frac{1}{p_0}}+2\lambda\beta.
\end{align}
Finally, by setting $\lambda=2$ we reach (\ref{eq.generalizedfoygel}).
}
\section{Proof of Lemma \ref{lemma.lip}}\label{proof.lemmalip}
	Define the functions 
	\begin{align}\label{eq.defs}
	&f_1(\bm{g}):={\rm dist}(\bm{g},\alpha\partial f(\bm{x})),\nonumber\\
	&f_2(\bm{g}):={\rm dist}(\bm{g},{\rm cone} (\partial f(\bm{x}))),\nonumber\\
	&h_1(\bm{g}):=f_1^2(\bm{g}),\nonumber\\
	&h_2(\bm{g}):=f_2^2(\bm{g}),
	\end{align}
	and the event 
	\begin{align}
	\mathcal{E}_1:=\Bigg\{f_2(\bm{g}) -\mathds{E}f_2(\bm{g})\le \zeta\Bigg\}.
	\end{align}
	Suppose that $\mathcal{E}$ and $\mathcal{E}_1$ hold. Then,
	\begin{align}\label{eq.liph1}
	&|h_1(\bm{g})-h_1(\bm{g}')|=|f_1(\bm{g})-f_1(\bm{g}')||f_1(\bm{g})+f_1(\bm{g}')|
	\le\nonumber\\
	&2\|\bm{g}-\bm{g}'\|_2(\zeta+\mathds{E}f_2(\bm{g})+2\lambda \beta),
	\end{align}
	where the second inequality comes from the fact that $f_1$ is $1$-Lipschitz function of $\bm{g}$. The last inequality is the result of $f_1(\bm{g})\le f_2(\bm{g})+2\lambda \beta$ (\ref{eq.generalized}). Now suppose that only $\mathcal{E}_1$ holds, Then, with the same reasoning, we have:
	\begin{align}\label{eq.liph2}
	&|h_2(\bm{g})-h_2(\bm{g}')|=|f_2(\bm{g})-f_2(\bm{g}')||f_2(\bm{g})+f_2(\bm{g}')|
	\le\nonumber\\
	&\|\bm{g}-\bm{g}'\|_2(|f_2(\bm{g})|+|f_2(\bm{g}')|)\le 2\|\bm{g}-\bm{g}'\|_2(\zeta+\mathds{E}f_2(\bm{g})),
	\end{align}
	\begin{align}\label{eq.help}
	&\mathds{P}\Big\{h_1-\mathds{E}[h_1]\le -\frac{r}{3}\bigg|\mathcal{E},\mathcal{E}_1\Big\}\le e^{-\frac{r^2}{72(\zeta+\mathds{E}f_2(\bm{g})+2\lambda \beta)^2}},\nonumber\\
	&\mathds{P}\Big\{h_2-\mathds{E}[h_2]\ge \frac{r}{3}\bigg|\mathcal{E}_1\Big\}\le e^{-\frac{r^2}{72(\zeta+\mathds{E}f_2(\bm{g}))^2}}.
	\end{align}
	Consequently,
	\begin{align}
	&\mathds{P}\{\phi_1(\bm{g})-\mathds{E}[\phi_1(\bm{g})]\le -r\}=\nonumber\\
	&\mathds{P}\Big\{h_1-\mathds{E}[h_1]-h_2+\mathds{E}[h_2]-4\lambda\beta f_2+4\lambda\beta\mathds{E}[f_2]\le -r\Big\} \nonumber\\
	&\le\mathds{P}\{h_1-\mathds{E}[h_1]\le - \frac{r}{3}\}+\mathds{P}\{h_2-\mathds{E}[h_2]\ge \frac{r}{3}\}+\nonumber\\
	&\mathds{P}\{f_2-\mathds{E}[f_2]\ge \frac{r}{12\lambda\beta}\}\le \mathds{P}\Big\{h_1-\mathds{E}[h_1]\le -\frac{r}{3}\Big|\mathcal{E}_1\Big\}\mathds{P}\{\mathcal{E}_1\}\nonumber\\
	&+\mathds{P}\Big\{h_1-\mathds{E}[h_1]\le -\frac{r}{3}\Big|\overline{\mathcal{E}}_1\Big\}\mathds{P}\{\overline{\mathcal{E}}_1\}+ \nonumber\\
	&\mathds{P}\Big\{h_2-\mathds{E}[h_2]\ge \frac{r}{3}\Big|\mathcal{E}_1\Big\}\mathds{P}\{\mathcal{E}_1\}+\nonumber\\
	&\mathds{P}\Big\{h_2-\mathds{E}[h_2]\ge \frac{r}{3}\Big|\overline{\mathcal{E}}_1\Big\}\mathds{P}\{\overline{\mathcal{E}}_1\}+\mathds{P}\{f_2-\mathds{E}[f_2]\ge \frac{r}{12\lambda\beta}\}\le\nonumber\\
	&e^{-\frac{r^2}{72(\zeta+\mathds{E}f_2(\bm{g})+2\lambda\beta)^2}}+2e^{-\frac{\lambda^2}{2}}+e^{-\frac{\zeta^2}{2}}+e^{-\frac{r^2}{72(\zeta+\mathds{E}f_2(\bm{g}))^2}}+e^{-\frac{\zeta^2}{2}}\nonumber\\
	&+e^{-\frac{r^2}{72\times 4\lambda^2\beta^2}}\le 3e^{-\frac{r^2}{72(\zeta+\mathds{E}f_2(\bm{g})+2\lambda\beta)^2}}+2e^{-\frac{\lambda^2}{2}}+2e^{-\frac{\zeta^2}{2}},
	\end{align}
	where in the third inequality, we used 
	\begin{align}
	&\mathds{P}\Big\{h_1-\mathds{E}[h_1]\le -\frac{r}{3}\Big|\mathcal{E}_1\Big\}=\mathds{P}\Big\{h_1-\mathds{E}[h_1]\le -\frac{r}{3}\Big|\mathcal{E}_1,\mathcal{E}\Big\}\mathds{P}\{\mathcal{E}\}\nonumber\\
	&+\mathds{P}\Big\{h_1-\mathds{E}[h_1]\le -\frac{r}{3}\Big|\mathcal{E}_1,\overline{\mathcal{E}}\Big\}\mathds{P}\{\overline{\mathcal{E}}\},
	\end{align}
	and (\ref{eq.help}). With a change of variable, we reach (\ref{eq.reform}).

\section{Proof of Proposition \ref{prop.l1analysis}}\label{proof.propl1ana}
The condition \eqref{eq.mycondition} for function $f=\|\bm{\Omega}\cdot\|_1$ can be stated as:
\begin{align}
\exists \bm{z}_0:~~\langle \bm{\Omega}^T\bm{w}-\bm{z}_0,\bm{z}_0\rangle=0~:\forall \bm{w}\in\partial \|\cdot\|_1(\bm{\Omega}\bm{x}).
\end{align}
By setting $\bm{w}={\rm sgn}(\bm{\Omega}\bm{x})+\bm{v}_{\overline{\mathcal{S}}}$ where $\|\bm{v}_{\overline{\mathcal{S}}}\|_{\infty}\le 1$, we have:
\begin{align}\label{eq.decom1}
&\langle \bm{\Omega}^T{\rm sgn}(\bm{\Omega}\bm{x})+\bm{\Omega}_{\overline{\mathcal{S}}}^T\widetilde{\bm{v}}_{\overline{\mathcal{S}}},\bm{z}_0\rangle=\|\bm{z}_0\|_2^2,\nonumber\\
&\forall \widetilde{\bm{v}}_{\overline{\mathcal{S}}}~\text{with}~\|\widetilde{\bm{v}}_{\overline{\mathcal{S}}}\|_{\infty}\le 1.
\end{align}
Since $\bm{z}_0$ and ${\rm sgn}(\bm{\Omega}\bm{x})$ are fixed and both $\widetilde{\bm{v}}_{\overline{\mathcal{S}}}$ and $-\widetilde{\bm{v}}_{\overline{\mathcal{S}}}$ satisfy $\|\widetilde{\bm{v}}_{\overline{\mathcal{S}}}\|_{\infty}\le 1$, it holds that:
\begin{align}\label{eq.decom2}
\langle \bm{\Omega}_{\overline{\mathcal{S}}}^T\widetilde{\bm{v}}_{\overline{\mathcal{S}}},\bm{z}_0\rangle =0.
\end{align}
The expressions \eqref{eq.decom1} and \eqref{eq.decom2} lead to:
\begin{align}
&\bm{z}_0\in{\rm null}(\bm{\Omega}_{\overline{\mathcal{S}}}),\label{eq.d1}\\
&\langle \bm{\Omega}^T{\rm sgn}(\bm{\Omega}\bm{x}),\bm{z}_0\rangle=\|\bm{z}_0\|_2^2.\label{eq.d2}
\end{align}
To satisfy \eqref{eq.d1} and \eqref{eq.d2} simultaneously, we choose the vector
\begin{align}
&{\bm{z}_0}=\tfrac{A_1}{A_2}\bm{P}_{{\rm null}(\bm{\Omega}_{\overline{\mathcal{S}}})}\bm{c}_0,
\end{align}
where
\begin{align*}
&A_1=\langle {\bm{\Omega}}^T {\rm sgn}({\bm{\Omega}}\bm{x}),\bm{P}_{{\rm null}(\bm{\Omega}_{\overline{\mathcal{S}}})}\bm{c}_0\rangle,\nonumber\\
&A_2=\|\bm{P}_{{\rm null}(\bm{\Omega}_{\overline{\mathcal{S}}})}\bm{c}_0\|_2^2,\nonumber\\
\end{align*}
and $\bm{c}_0$ is an arbitrary vector.
By choosing $\bm{c}_0=\bm{\Omega}^T {\rm sgn}(\bm{\Omega x})$, we have:
\begin{align}
\bm{z}_0=\bm{P}_{{\rm null}(\bm{\Omega}_{\overline{\mathcal{S}}})}\bm{\Omega}^T{\rm sgn}(\bm{\Omega x})
\end{align}


\section{Numerical computation of the Gaussian width}\label{sec.TrueGaussianwidth}
The Gaussian width $\omega(\mathcal{D}(f,\bm{x})\cap \mathbb{B}^n)$ plays a key role in our proposed error estimate in Theorem \ref{thm.maintheorem}. In this appendix, we explain how to compute this quantity  numerically. This approach is adapted from \cite[Section B.2]{genzel2017ell}. Recall that this quantity is defined as
\begin{align}
\omega(\mathcal{D}(f,\bm{x})\cap \mathbb{B}^n):=\mathds{E}\sup_{\substack{\bm{y}\in \mathcal{D}(f,\bm{x})\\\|\bm{y}\|_2\le 1}}\langle \bm{y},\bm{g}\rangle.
\end{align}
By choosing a sufficiently small $t$, e.g. $t=0.01$ in \eqref{eq.descent cone} (the rationale of this choice is discussed in \cite[Section B.2]{genzel2017ell}), the expression inside $\mathds{E}$ can be simplified as the simple convex program:
\begin{align}
\omega(\mathcal{D}(f,\bm{x})\cap \mathbb{B}^n)\approx\mathds{E}\sup_{\substack{f(\bm{x}+t\bm{y})\le f(\bm{x})\\\|\bm{y}\|_2\le 1}}\langle \bm{y},\bm{g}\rangle,
\end{align}
that can be solved using the CVX package \cite{cvx}. 

\section{The weak decomposability condition for $\ell_1$-analysis}\label{sec.l1ana_fail}
With a counterexample, we show that the mentioned decomposability condition does not hold in general. 
Let $\bm{\Omega}_{p\times n}$ be a tall analysis operator in general position and $\bm{x}_{n\times 1}$ be an analysis-sparse vector such that $\mathcal{S}={\rm supp}(\bm{\Omega}\bm{x})$ with $|\mathcal{S}|\geq p-n$ (see \cite[Section 2.1]{nam2013cosparse}). 
For the weak decomposability condition of \cite{foygel2014corrupted} to hold for $(\bm{\Omega},\mathcal{S},\bm{x})$, we shall have that
\begin{align}
&\exists \,\bm{w}_0\in\partial \|\cdot\|_{1}(\bm{\Omega}\bm{x}),~~\nonumber\\
&\forall \bm{w}\in\partial \|\cdot\|_1(\bm{\Omega}\bm{x}):~~~\langle \bm{\Omega}^T(\bm{w}-\bm{w}_0) ~,~\bm{\Omega}^T\bm{w}_0\rangle=0.
\end{align}
In particular, we can set $\bm{w}={\rm sgn}({\bm{\Omega} \bm{x}})+{\bm v}_{\overline{\mathcal{S}}}$, where ${\bm v}_{p\times 1}$ is an arbitrary vector with $\|{\bm v}\|_{\infty}\leq 1$. Since $(\bm{\Omega}\bm{ x})_{\overline{\mathcal{S}}}={\bm 0}_{p\times 1}$, we can write that
\begin{align}
&\forall \,\bm{v},~\|\bm{v}\|_{\infty}\le 1:~~~\nonumber\\
&\langle \bm{\Omega}^T\,{\rm sgn}(\bm{\Omega}\bm{x})+\bm{\Omega}_{\overline{\mathcal{S}}}^T\,\widetilde{\bm{v}}_{\overline{\mathcal{S}}} ~,~\bm{\Omega}^T\bm{w}_0\rangle=\|\bm{\Omega}^T\bm{w}_0\|_2^2.
\end{align}
By setting $\bm v =\bm 0$ and applying the result for general $\bm v$, we obtain
\begin{align}
\langle \bm{\Omega}_{\overline{\mathcal{S}}}^T \widetilde{\bm{v}}_{\overline{\mathcal{S}}} \,,\, \bm{\Omega}^T\bm{w}_0\rangle =0,
\end{align}
for arbitrary $\bm{v}$ with $\|\bm{v}\|_{\infty}\le 1$, or equivalently, for arbitrary $\bm{v}$. This implies that
\begin{align}
\langle \widetilde{\bm{v}}_{\overline{\mathcal{S}}} \,,\, \bm{\Omega}_{\overline{\mathcal{S}}} \, \bm{\Omega}^T\bm{w}_0\rangle =0,
\end{align}
or equivalently
\begin{align}
\bm{\Omega}_{\overline{\mathcal{S}}}\,\bm{\Omega}^T\bm{w}_0 =0.
\end{align}
As $\bm{w}_0\in\partial \|\cdot\|_{1}(\bm{\Omega}\bm{x})$, we know that $\bm{w}_0={\rm sgn}(\bm{\Omega x})+{\bm{v}_0}_{\overline{\mathcal{S}}}$ for some $\|\bm{v}_0\|_{\infty}\leq 1$.Therefore,
\begin{align}
\bm{\Omega}_{\overline{\mathcal{S}}} \, \bm{\Omega}_{\overline{\mathcal{S}}}^T \, {\widetilde{\bm{v}_0}}_{\overline{\mathcal{S}}}
= -\bm{\Omega}_{\overline{\mathcal{S}}} \, \bm{\Omega}^T \, {\rm sgn}(\bm{\Omega} \, \bm{x})
= -\bm{\Omega}_{\overline{\mathcal{S}}} \, \bm{\Omega}_{\mathcal{S}}^T \, {\rm sgn}(\bm{\Omega}_{\mathcal{S}} \,\bm{x}).
\end{align}
Because $\bm{\Omega}$ is in general position, $\bm{\Omega}_{\overline{\mathcal{S}}} \, \bm{\Omega}_{\overline{\mathcal{S}}}^T$ is invertible and we can express ${\widetilde{\bm{v}_0}}_{\overline{\mathcal{S}}}$ as
\begin{align}
\widetilde{\bm{v}_0}_{\overline{\mathcal{S}}}
= - \big(\bm{\Omega}_{\overline{\mathcal{S}}} \, \bm{\Omega}_{\overline{\mathcal{S}}}^T\big)^{-1}\bm{\Omega}_{\overline{\mathcal{S}}} \, \bm{\Omega}_{\mathcal{S}}^T \, {\rm sgn}(\bm{\Omega}_{\mathcal{S}} \,\bm{x}).
\end{align}
Now, the contradiction comes from the fact that the entries $\bm{v}_0$ in the above equation are not necessarily confined to the interval $[-1,1]$. We show this by a numerical example:
%
\begin{align}
\bm{x}=\begin{bmatrix}
-.4472\\
.8944
\end{bmatrix}, \bm{\Omega}=\begin{bmatrix}
1&1\\
2&1\\
1&2
\end{bmatrix}
,\mathcal{S}=\{1,3\},\overline{\mathcal{S}}=\{2\}.
\end{align}
For the latter signal, it holds that
\begin{align}
\widetilde{v_0}_{\overline{\mathcal{S}}}=-1.4,
\end{align}
which obviously contradicts the constraint $\|{v_0}_{\overline{\mathcal{S}}}\|_{\infty}\le 1$.

\section{Asymptotic behavior of \texorpdfstring{$\beta$}{TEXT} when  \texorpdfstring{$\bm{\Omega}$}{TEXT} is Gaussian ensemble}\label{beta_gaussian}
 
 We find an upper-bound on $\beta$ when the analysis operator is a Gaussian ensemble (whether fat or tall). Recall that an upper-bound for $\beta$ is obtained in \eqref{eq.upper_on_beta} as follows:
\begin{align}
\beta\le\frac{\|\bm{\Omega}_{\mathcal{S}}^T{\rm sgn}(\bm{\Omega}_{\mathcal{S}}\bm{x})\|_2}{\|\bm{P}_{{\rm null}(\bm{\Omega}_{\overline{\mathcal{S}}})}\bm{\Omega}_{\mathcal{S}}^T{\rm sgn}(\bm{\Omega}_{\mathcal{S}}\bm{x})\|_2}.
\end{align}
Since $\bm{\Omega}_{\mathcal{S}}$ is statistically independent of $\bm{\Omega}_{\overline{\mathcal{S}}}$, ${\bm P}_{{\rm null}({\bm \Omega}_{\overline{S}})}$ defines a projection onto an $n-|\overline{\mathcal{S}}|$ subspace that is independently and uniformly oriented with respect to $\bm{v}=\frac{{\bm \Omega}_{\mathcal{S}}^T \,{\rm sgn} (\, {\bm \Omega}_{\mathcal{S}} \, {\bm x})}{\|{\bm \Omega}_{\mathcal{S}}^T \,{\rm sgn} (\, {\bm \Omega}_{\mathcal{S}} \, {\bm x})\|_2}$. {\color{\change} In addition
\begin{align}\label{eq:unit_v}
\beta \le \frac{1}{\|\bm{P}_{{\rm null}(\bm{\Omega}_{\overline{\mathcal{S}}})} \bm{v}\|_2} .
\end{align}
As the relative orientation of $\bm{v}$ with respect to ${\rm null}(\bm{\Omega}_{\overline{\mathcal{S}}})$ determines the upper-bound for $\beta$, 
 we can fix $\bm{v}$ on the unit sphere and randomly rotate ${\rm null}(\bm{\Omega}_{\overline{\mathcal{S}}})$ with a uniform Haar measure. Equivalently, we can fix ${\rm null}(\bm{\Omega}_{\overline{\mathcal{S}}})$ and randomly select $\bm{v}$ with a uniform distribution on the unit sphere. In fact, the choice $\bm{v}=\frac{\bm{g}}{\|\bm{g}\|_2}$ where $\bm{g}$ is an i.i.d. random vector with standard normal distribution independent of ${\rm null}(\bm{\Omega}_{\overline{\mathcal{S}}})$ fulfills this requirement. With this choice, we can rewrite
 the upper-bound on $\beta$ as
\begin{align}
\beta\le \frac{\|\bm{g}\|_2}{\|\bm{P}_{{\rm null}(\bm{\Omega}_{\overline{\mathcal{S}}})}\bm{g}\|_2}.
\end{align}
%
%
It is straightforward to check that $\|\bm{P}_{{\rm null}(\bm{\Omega}_{\overline{\mathcal{S}}})}\cdot\|_2$ and $\|\cdot\|_2$ are $1$-Lipschitz functions.
Therefore, in high dimensions, we know that $\|\bm{P}_{{\rm null}(\bm{\Omega}_{\overline{\mathcal{S}}})}\bm{g}\|_2$ and $\|\bm{g}\|_2$ are concentrated around $\mathds{E}\|\bm{P}_{{\rm null}(\bm{\Omega}_{\overline{\mathcal{S}}})}\bm{g}\|_2$ and $\mathds{E}\|\bm{g}\|_2$, respectively. Also, by \cite[Corollary 3.2]{ledoux2001concentration} and \cite[Proposition 8.1]{foucart2013mathematical}, it holds that \begin{align}\label{eq.cons_bounds1}
&\mathds{E}\|\bm{P}_{{\rm null}(\bm{\Omega}_{\overline{\mathcal{S}}})}\bm{g}\|_2\ge \sqrt{\mathds{E}\|\bm{P}_{{\rm null}(\bm{\Omega}_{\overline{\mathcal{S}}})}\bm{g}\|_2^2-1},\nonumber\\
&\mathds{E}\|\bm{g}\|_2\le \sqrt{n}.
\end{align}
Let $[\bm{u}_1, \dots, \bm{u}_{n-|\overline{\mathcal{S}}|}]$ be a basis for ${\rm null}(\bm{\Omega}_{\overline{\mathcal{S}}})$. Then, we have
\begin{align}\label{eq:beta_E}
&\mathds{E}\|\bm{P}_{{\rm null}(\bm{\Omega}_{\overline{\mathcal{S}}})}\bm{g}\|_2^2=\sum_{i=1}^{n-|\overline{\mathcal{S}}|}\mathds{E}|\langle \bm{u}_i, \bm{g}\rangle|^2=\sum_{i=1}^{n-|\overline{\mathcal{S}}|}\bm{u}_i^T\mathds{E}\bm{g}\bm{g}^T\bm{u}_i.
\end{align}
Because $\mathds{E}\bm{g}\bm{g}^{\rm H}=\bm{I}$,
we can now simplify \eqref{eq:beta_E} as
\begin{align}\label{eq:beta_E1}
&\mathds{E}\|\bm{P}_{{\rm null}(\bm{\Omega}_{\overline{\mathcal{S}}})}\bm{g}\|_2^2
=\sum_{i=1}^{n-|\overline{\mathcal{S}}|} \underbrace{\bm{u}_i^T\bm{u}_i}_{1} =n-|\overline{\mathcal{S}}|.
\end{align}
As a result, due to \eqref{eq.cons_bounds1} and \eqref{eq:beta_E1}, we reach

$$\beta\le \frac{\sqrt{n}}{\sqrt{n-|\overline{\mathcal{S}}|-1}}.$$}
%

\section{Analysis operators with orthogonal rows}\label{proof.beta_one}
In this section, we consider a fat analysis operator $\bm{\Omega}\in\mathbb{R}^{p\times n}$ that is constructed via the recipe in Item \ref{item.random1} of Section \ref{section.simulation}. The result, however,  holds for all analysis operators for which $\bm{\Omega}\bm{\Omega}^T$ is diagonal. 

When $\bm{\Omega}$ in \eqref{eq.random1operators} is fat with full row-rank, we have $r=p$ and $\bm{\Omega}$ can be expressed as
\begin{align}
\bm{\Omega}=\bm{D}_{p\times p}\bm{U}_{p\times p}\bm{V}_{n\times p}^{\rm T}.
\end{align}
Using MATLAB matrix notations, we have:
\begin{align}
&\bm{\Omega}_{\overline{\mathcal{S}}}=\bm{D}(\overline{\mathcal{S}},\overline{\mathcal{S}})\bm{U}(\overline{\mathcal{S}},[p])\bm{V}([n],[p])^{\rm T},\nonumber\\
&\bm{\Omega}_{\mathcal{S}}=\bm{D}({\mathcal{S}},{\mathcal{S}})\bm{U}({\mathcal{S}},[p])\bm{V}([n],[p])^{\rm T}.
\end{align}
As a consequence, it holds that
\begin{align}
\bm{\Omega}_{\overline{\mathcal{S}}}\bm{\Omega}_{\mathcal{S}}^{\rm T}=\bm{D}(\overline{\mathcal{S}},\overline{\mathcal{S}})\bm{U}(\overline{\mathcal{S}},[p])\bm{U}^{\rm T}({\mathcal{S}},[p])\bm{D}({\mathcal{S}},{\mathcal{S}}).
\end{align}
Since $\bm{U}(\overline{\mathcal{S}},[p])\bm{U}^{\rm T}({\mathcal{S}},[p])$ is a submatrix of $\bm{U}\bm{U}^T=\bm{I}_p$, we have that
\begin{align}
\bm{\Omega}_{\overline{\mathcal{S}}}\bm{\Omega}_{\mathcal{S}}^{\rm T}=\bm{0}.
\end{align}
Thus, the denominator of $\beta$ becomes
\begin{align}\label{eq.beta_denom}
\|\bm{P}_{{\rm null}(\bm{\Omega}_{\overline{\mathcal{S}}})}\bm{\Omega}^T{\rm sgn}(\bm{\Omega x})\|_2 &= \|(\bm{I}_n-\bm{\Omega}_{\overline{\mathcal{S}}}^\dagger\bm{\Omega}_{\overline{\mathcal{S}}})\bm{\Omega}_{\mathcal{S}}^T{\rm sgn}(\bm{\Omega}_{\mathcal{S}} \bm{x}) \|_2\nonumber\\
&=\|\bm{\Omega}_{\mathcal{S}}^T{\rm sgn}(\bm{\Omega}_{\mathcal{S}}\bm{ x}) \|_2.
\end{align}
By checking the gradient of the cost in the optimization in the definition of $\beta$ (numerator), we can check that
\begin{align}
\bm{z}_{\overline{\mathcal{S}}}=-(\bm{\Omega}_{\overline{\mathcal{S}}}\bm{\Omega}_{\overline{\mathcal{S}}}^T)^{-1}\bm{\Omega}_{\overline{\mathcal{S}}}\bm{\Omega}^T_{\mathcal{S}}{\rm sgn}(\bm{\Omega}_{\mathcal{S}}\bm{x})=\bm{0}
\end{align}
is the unique minimizer (the gradient of the cost is zero at $\bm{z}_{\overline{\mathcal{S}}}$, and $\bm{z}_{\overline{\mathcal{S}}}$ satisfies the constraints). Also, the minimum value of the cost (value of the numerator) becomes  $\|\bm{\Omega}^T{\rm sgn}(\bm{\Omega x})\|_2$ with this choice of $\bm{z}_{\overline{\mathcal{S}}}$. This reveals that the numerator of $\beta$ equals its denominator, i.e., $\beta=1$.
%
\section*{Acknowledgment}
The authors thank the anonymous reviewers for valuable comments and suggestions to improve the quality of the paper. S.Daei also wishes to thank Mohammad Ali Hoseini Nasab for fruitful discussions. 
\ifCLASSOPTIONcaptionsoff
  \newpage
\fi

\bibliographystyle{ieeetr}
\bibliography{mypaperbibe}
\end{document}